\documentstyle[11pt,fps]{article}
\newcommand{\Z}{Z \!\!\! Z}

\newcommand{\D}{{\cal D}}
\newcommand{\vp}{\varphi}
\newcommand{\sla}{\slash \!\!\!}
\newcommand{\unit}{1 \!\! 1}
\makeatletter
\@addtoreset{equation}{section}
\makeatother

\begin{document}
\vspace*{1cm}
\begin{center}

{\LARGE Perfect Scalars on the Lattice}

\vspace*{1cm}

W. Bietenholz
\vspace*{1cm}

NORDITA \\
Blegdamsvej 17 \\
DK-2100 Copenhagen \O , Denmark

\vspace*{1cm}

Preprint NORDITA-1999/69-HE

\end{center}

\vspace*{15mm}

We perform renormalization group transformations to
construct optimally local perfect lattice actions for 
free scalar fields of any mass. Their couplings decay 
exponentially. The spectrum is identical to the continuum 
spectrum, while thermodynamic quantities
have tiny lattice artifacts. To make such actions
applicable in simulations, we truncate the couplings to a unit 
hypercube and observe that spectrum and thermodynamics
are still drastically improved compared to the standard
lattice action. We show how preconditioning techniques
can be applied successfully to this type of action.
We also consider a number of variants of the perfect lattice
action, such as the use of an anisotropic or triangular
lattice, and modifications of the renormalization group
transformations motivated by wavelets. Along the way
we illuminate the consistent treatment of gauge fields, and
we find a new fermionic fixed point action with attractive 
properties.

\newpage

\section{Introduction}

{\em Renormalization group transformations} (RGTs) are an
important operation, both, in field theory and in statistical 
mechanics. They allow to vary the cutoff of a system without 
altering its physical contents. In particular RGT fixed points
and their vicinity are of interest, both, conceptually and for 
practical purposes.

If a system is defined on a (Euclidean) lattice, then
RGTs can be performed by block variable transformations
and a subsequent rescaling \cite{WiKo}. In particular for
the simple system of a free scalar field, it has been shown
a long time ago that such block variable RGTs -- for instance 
with blocking factor 2 -- can be iterated an infinite number of
times in momentum space \cite{BeWi}. For suitably chosen
RGT parameters and massless scalars, one arrives at a
finite {\em fixed point action} (FPA). There the observables
are independent of the lattice spacing -- which is
an inverse UV cutoff -- so that we can
represent exact continuum physics on the lattice.

In each transformation step the average of the variables 
in one block of 
the fine lattice is related to the corresponding block variable
on the coarse lattice.
This has often been implemented by a $\delta$ function.
However, the $\delta$ function is just one possibility in a 
class of implementations, which keep
the partition function and all expectation values invariant.
In Ref.\ \cite{BeWi} a generalization was suggested,
where this relation is smoothly implemented by a Gaussian,
and it was observed that in this way the locality of the
FPA can be improved. 
\footnote{Also the scalar FPA of the $\delta$ function RGT is 
local in the sense of an exponential decay of the couplings
(in contrast to the fermionic FPA), but for certain Gaussian 
RGTs this decay becomes even faster, see Section 2.}
This is very important for practical
purposes, because one ultimately has to truncate the
couplings to short distances, and to hope that this 
truncation does not affect the perfect properties too much.
If one sends the blocking factor to infinity, then
one single RGT leads to the FPA \cite{CG,Schwing}.
This amounts to a technique that we call ``blocking
from the continuum'': one starts from a continuum
theory and defines lattice variables by relating
each of them to the Riemann integral over a lattice
cell.

A similar construction has been realized for free
fermions with a $\delta$ function RGT \cite{Hambu} and
a generalized RGT \cite{UJW}. In that context the FPA
also yielded new insight into the fermion doubling problem.
By a somewhat more complicated RGT one obtains
a FPA for staggered fermions too \cite{Mai}.
Also here we can optimize locality by using a smooth
RGT \cite{Dallas} and the result can be reproduced again 
by blocking from the continuum \cite{Dilg,stag}.
Moreover, blocking from the continuum also leads to
a fixed point for free non-compact gauge fields \cite{QuaGlu}.

FPAs have the remarkable property that a system is regularised
but not contaminated by any cutoff artifacts.
Of course it is a dream to construct such
a regularisation also for interacting theories at arbitrary
correlation length $\xi$ (FPAs only exist at $\xi = \infty$).
Indeed, general
arguments show that such {\em ``perfect lattice actions''} do exist 
even at finite correlation length \cite{WiKo}, but they are
hard to construct.  By blocking from the
continuum this can be achieved at least perturbatively, 
and the perfect action was constructed explicitly
to first order in the fermion-gauge coupling for the
Schwinger model \cite{Schwing} and for QCD \cite{QuaGlu,StL,Org}.
For the anharmonic oscillator, the first order perturbatively
perfect action has been constructed and simulated in order to
test its scaling and asymptotic scaling behavior \cite{BS}.

This approach to fight artifacts due to the finite
lattice spacing -- which are the worst systematic errors
in many Monte Carlo simulations -- has picked up new momentum
since the appearance of Ref.\ \cite{Has}. There the authors
noticed that for asymptotically free theories the construction
of a FPA (with rescaling of the weakly relevant coupling)
is a classical field theory problem. 
\footnote{Alternatively there are many attempts to use
real space blocking RGTs (without classical approximation).
For recent work on scalars resp.\ pure $SU(3)$ gauge fields,
see Ref.\ \cite{RSRG}. A limitation there is that one can hardly
implement the blocking constraint in a way different from the
$\delta$ function.}
Hence it can be solved
by minimization instead of (numerical) functional integrals,
which simplifies the task enormously. Moreover the authors
suggest to use the FPA, which is a classically perfect
action, even at moderate correlation length, and they
demonstrated for the 2d $O(3)$ model that this can be
a very successful approximation of a perfect lattice action.
The construction can be performed non-perturbatively, by numerical
inverse blocking RGTs. Of course for complicated theories like QCD,
or pure $SU(3)$ gauge theory \cite{tdg},
even this approximation is tedious.
In addition, problems related to the parameterization and truncation of
such quasi-perfect actions are difficult to handle, but crucial for
potential applications. With this respect it is advisable 
to start from a simple situation and study there, which procedures are
promising. The present paper is a contribution to that issue. 
A synopsis of our results was anticipated in Ref.\ \cite{Lat97}.\\

In this paper we deal with scalar fields.
In analogy to the previously studied fermions, the free
particles provide an improved formulation, which is promising
also in the presence of interactions.
In Section 2 we derive perfect lattice actions for
free lattice fields by blocking from the continuum and
we optimize their locality. In Section 3 we truncate the
couplings to a unit hypercube and consider the effect
on the dispersion relation. In Section 4 we illustrate the improved
thermodynamic scaling properties of this {\em ``hypercube scalar''}.
For the evaluation of the scalar matrix -- analogous to
the fermion matrix -- we can still use
preconditioning in the hypercubic case, as we show in Section 5.
In Section 6 we discuss
perfect actions on anisotropic and triangular lattices.
As a further alternative we consider in Section 7 a 
class of new blocking prescriptions, which is motivated by
the wavelet resp.\ B-spline formalism. The most promising new
variant is also applied to fermions (Appendix B).
Section 8 contains our conclusions and an outlook on applications.

\section{Perfect lattice scalar actions}

Let us start from a scalar field $\phi$ and its
action $S[\phi ]$ on a hypercubic
lattice in $d$ dimensional Euclidean space.
The Kadanoff transformation to a coarser
lattice can be written as \cite{Kada}
\begin{equation}
e^{-S'[\phi ']} = \int \D \phi \ K[\phi ',\phi ] e^{-S[\phi ]}
\end{equation}
where $S'[\phi ']$ is the action of the new scalar
field $\phi'$ on the coarse lattice. We integrate over
the lattice measure $\D \phi = \int \prod_{x} d \phi_{x}$.
The kernel $K[\phi ',\phi ]$ has to be chosen such that the
partition function and its derivatives remain invariant
under this transformation (up to a possible constant factor).
This can be achieved by the requirement
\begin{equation} \label{RGTcon}
\int \D \phi' \ K[\phi', \phi ] = {\rm const} .
\end{equation}

The simplest and most popular
choice is 
\begin{equation} \label{deltatrafo}
K[\phi ',\phi ] = \prod_{x'} \delta (\phi '_{x'} - A_{x'} [\phi ]).
\end{equation}
Here $x'$ runs over the coarse lattice, and the symbol
$A_{x'}[\phi ]$ is an average taken in the block of the
fine lattice, which is attached to $x'$. For instance, one
may just take the arithmetic mean value together
with a rescaling factor,
\begin{equation} \label{blocksum}
A_{x'} [ \phi ] = \frac{b_{n}}{n^{d}} \sum_{x \in x'} \phi_{x} .
\end{equation}
This sum extends over all the $n^{d}$ fine lattice points 
in the block
associated with $x'$. After the transformation we express
all quantities in units of the coarse lattice. The factor
$b_{n}$ is used to neutralize this rescaling so 
that we can arrive at a finite FPA.
Hence its value is determined by the dimension of the scalar field,
\begin{equation}
b_{n} = n^{(d-2)/2} .
\end{equation}

However, the choice (\ref{deltatrafo}) is by no means unique.
A class of generalizations has been considered by Bell and
Wilson, where the $\delta$ function is replaced by a
Gaussian \cite{BeWi},
\begin{equation} \label{gautrafo}
K[\phi ' ,\phi ] = \exp \Big\{ - \frac{2}{\alpha} \sum_{x'}[\phi'_{x'}-
A_{x'} \phi ]^{2} \Big\} \  , \qquad (\alpha \geq 0 ) . 
\end{equation}
In the limit $\alpha \to 0$ we return to the $\delta $ RGT
(\ref{deltatrafo}), but the condition (\ref{RGTcon})
holds for all positive $\alpha$.\\

In Ref.\ \cite{BeWi} the fixed point for a massless scalar
particle has been computed by an infinite number of
iterations of such block factor 2 RGTs. The iteration starts
from the standard lattice action in momentum space,
\begin{equation} \label{stadact}
S [\phi ] = \frac{1}{(2\pi )^{d}} \int d^{d}k \ \phi (-k)
\frac{1}{2} \hat k^{2} \phi (k) \ , 
\qquad (\hat k_{\mu} := 2 \sin \frac{k_{\mu}}{2}),
\end{equation}
where we set the lattice spacing equal to 1.

However, instead of this tedious iteration we can also send 
the blocking factor $n \to \infty$, so we obtain a FPA
by performing only one RGT \cite{CG,Schwing}.
Since we call the spacing of the final lattice 1 again,
the original lattice variables move closer and closer
together (in the final units) as $n$ increases.
Finally the sum in condition (\ref{RGTcon}) turns into a
Riemann integral,
\begin{equation} \label{cellint}
A_{x} \vp = \int_{C_{x}} d^{d}y \ \vp (y) .
\end{equation}
Here we do not start from a field $\phi_{x}$ on a fine lattice 
any longer, but from a continuum field $\vp (y)$. We
integrate over hypercubic unit cells $C_{x}$ with centers $x$,
and relate those integrals to the new lattice variables
$\phi_{x}$. Also the action on the fine lattice is replaced
by the continuum action. Now the RGT step requires
a continuum path integral, so it can only be computed
in certain cases.
Of course this limit $n\to \infty $ can also
be taken for the generalized form (\ref{gautrafo}).\\

We first reproduce this calculation in its short-cut form, 
including also an arbitrary mass $m$ of the scalar field.
There is no FPA at $m>0$, but we construct
an entire renormalized trajectory, i.e.\ a curve
of perfect lattice actions, for free scalars of any mass. It
is parameterized by $m$ and crosses the critical surface in
a fixed point at $m=0$.

The lattice action is given by
\begin{eqnarray}
e^{-S[\phi ]} &=& \int \D \phi \D \sigma \exp \Big\{
- \frac{1}{(2\pi )^{d}} \int_{B} d^{d}k \times \nonumber \\
&& \Big[ \frac{1}{2} \sum_{l \in \Z^{d}} \vp (-k-2\pi l)
[(k+2\pi l)^{2}+m^{2}] \vp (k+2\pi l) \nonumber \\
&& + i \sigma (-k) [ \phi (k) - \sum_{l\in \Z^{d}}
\vp (k+2\pi l) \Pi (k+2\pi l) ]        \label{RGT}
+ \frac{\alpha}{2} \sigma (-k) \sigma (k) \Big] \Big\},
\end{eqnarray}
where $B = \ ]-\pi , \pi]^{d}$ is the (first) Brillouin zone.
We have introduced an auxiliary lattice scalar variable
$\sigma$, and the function $\Pi$ is defined as
\begin{equation}
\Pi (k) = \prod_{\mu = 1}^{d} \frac{\hat k_{\mu}}{k_{\mu}} .
\end{equation}

We can write eq.\ (\ref{cellint}) as a convolution,
\begin{equation} \label{cellintg}
A_{x} \phi = \int d^{d}y \ f(x-y) \vp (y) \ , \quad
f(u) = \Big\{ \begin{array}{cc} 1 & \vert u_{\mu} \vert 
\leq \frac{1}{2}, \ \mu = 1, \dots ,d \\
0 & {\rm otherwise} \end{array} ,
\end{equation}
and $\Pi (k)$ is the Fourier transform
of $f(u)$. Instead of the piece-wise constant function
used here one may also insert other functions $f(u)$, 
if they fall off sufficiently fast. Some options will 
be discussed in Section 7.

We denote the continuum propagator as
\begin{equation}
\Delta (k) := \frac{1}{k^{2}+m^{2}} ,
\end{equation}
substitute
\begin{equation}
\tilde \vp (k+2\pi l) = \vp (k+2\pi l) -i\sigma (k)
\Delta(k+2\pi l) \Pi (k+2\pi l),
\end{equation}
and carry out the Gaussian integration over $\tilde \vp$:
\begin{eqnarray}
e^{-S[\phi ]} &=& \int \D \sigma \exp \Big\{ \frac{1}{(2\pi )^{d}}
\int_{B} d^{d}k [i\sigma(-k)\phi(k) - \frac{1}{2} \sigma(-k)
G(k) \sigma(k) ] \Big\}  , \nonumber \\
G(k) & := & \sum_{l \in \Z^{d}} \Delta (k+2\pi l) \Pi^{2}
(k+2\pi l) + \alpha . \label{perfprop}
\end{eqnarray}
By further substituting
\begin{equation}
\tilde \sigma (k) = \sigma (k) - i G(k)^{-1} \phi(k)
\end{equation}
and integrating $\tilde \sigma$ we arrive at
\begin{equation} \label{PA}
S[\phi ] = \frac{1}{(2\pi )^{d}} \int_{B} d^{d}k \ \frac{1}{2}
\phi (-k) G(k)^{-1} \phi (k) ,
\end{equation}
which shows that $G(k)$ is the perfect free lattice propagator.
Note that this scalar perfect lattice
action is also the renormalized trajectory 
of the $O(N)$ non-linear $\sigma$ model in the limit $N\to \infty$ 
(up to a factor $N$) \cite{HS,Anton},
in analogy to the large $N$ limit of the Gross Neveu model \cite{BFW}.
The sum over $l \in \Z^{d}$ converges in any dimension
thanks to the $\Pi $ function, but it can be computed
analytically only in $d=1$. If we now specify the RGT
``smearing parameter'' (because it smears the $\delta $
function to a Gaussian) to be
\begin{equation} \label{smear}
\alpha = \bar \alpha (m) := \frac{\sinh m - m}{m^{3}},
\end{equation}
then the perfect action in $d=1$ becomes ultralocal,
\begin{equation}
G_{d=1}(k) = \frac{\sinh m \cdot \hat m^{2}}{m^{3}} \frac{1}
{\hat k^{2} + \hat m^{2}} \ , \quad 
\hat m := 2 \sinh \frac{m}{2}.
\end{equation}
In this case, the couplings are even restricted to nearest 
neighbors. In particular the 1d fixed point propagator
(defined at $m=0$, where $\bar \alpha = 1/6$) is identical 
to the standard lattice lattice propagator $1/\hat k^{2}$. 
This is obviously
the choice for $\alpha$, which optimizes the locality in $d=1$.
For a general blocking factor $n$, the 1d perfect
propagator turns ultralocal with
\begin{equation} \label{aln}
\bar \alpha_{n}(m) := \bar \alpha (m) \ \Big( 1 - \frac{1}{n^{2}} \Big) .
\end{equation}

In higher dimensions -- that is, in field theory --
locality of the perfect propagator can only
be achieved in the sense of an exponential decay. For fermions
the RGT parameters which provide ultralocality in $d=1$ 
have been used successfully also in higher dimensions. It turned
that they still yield extremely local perfect actions in $d>1$
\cite{UJW,Dallas,QuaGlu,stag}. For free gauge fields, a similar
(but somewhat more complicated) smearing term leads to
ultralocality (the standard plaquette action)
in $d=2$ and extreme locality in $d=4$
\cite{QuaGlu,stag}. Hence we are guided to apply the same
strategy to scalar fields and use $\alpha = \bar \alpha$ as given
in eq.\ (\ref{smear}) in any dimension.

Bell and Wilson compared numerically various smearing
parameters in $d=3$, and in the fixed point they found optimal
locality at $1/\alpha_{2} \simeq 8$,
which agrees with the value resulting from our analytical method.

The perfectness of the action (\ref{PA}) can be verified by
an explicit additional block factor $n$ transformation
($n< \infty$). This is demonstrated in Appendix A.
The result is that a block factor $n$ RGT applied on
$G(k)$ at mass $m$ reproduces $G(k)$, now at mass $n\cdot m$.
This is exactly the expected behavior on the renormalized 
trajectory. This calculation also justifies eq.\ (\ref{aln})
as a generalization of $\bar \alpha = \bar \alpha_{\infty}$ given in
eq.\ (\ref{smear}).\\

Let us consider the spectrum of the perfect lattice action 
(\ref{PA}). We denote $k=(\vec k , k_{d})$ and identify the energy as 
$E=i(k_{d}+2\pi l_{d})$. Thus the spectrum is given by
\begin{equation} \label{spec}
E^{2} = (\vec k + 2\pi \vec l )^{2} + m^{2} .
\end{equation}
There is a branch for each $\vec l$. The branch for $\vec l = \vec 0$
is the {\em exact} continuum spectrum. In this sense,
the full, continuous Poincar\'{e} invariance is present
in the perfect lattice action. We emphasize that the form of the
lattice action itself does not reveal this property: for instance,
the hypercubic structure of the lattice is visible in it,
and perfect actions on non hypercubic lattices have
different forms, see Section 6 for examples. However,
physical observables do have the full continuum symmetries,
without any artifacts due to the finite lattice spacing.

The higher branches in the spectrum are required for perfection,
since the lattice imposes $2\pi$ periodicity on the momenta.\\

We now address the question of {\em locality}. For massless
Wilson-like fermions and a $\delta$ function RGT one obtains a nonlocal
fixed point. Its couplings do not decay exponentially, but only
$\propto r^{1-d}$ \cite{Hambu}. This is the way a contradiction
with the Nielsen-Ninomiya theorem is avoided. 
\footnote{For any $\alpha_{f}>0$ the fermionic FPA turns local
\cite{UJW}, but then the chiral symmetry of the lattice action
is explicitly broken, avoiding again a contradiction with the
Nielsen-Ninomiya No-Go theorem. However, this only means that the chiral
symmetry is not manifest in the lattice action. In physical 
observables it is still present \cite{GW,Melb}, similar to the 
Poincar\'{e} invariance, which is present in the spectrum, although
not manifest in the action.}
However, there is no theorem forbidding the scalar $\delta $ function 
FPA to be local. We analyze this issue in $d=1$.

A perfect lattice fermion propagator -- constructed in analogy
to eq.\ (\ref{RGT}) -- reads \cite{QuaGlu}
\begin{equation}
G^{f}(k) = \sum_{l \in \Z^{d}} \frac{\Pi (k+2\pi l)^{2}}
{i (\sla k + 2\pi \sla l ) +m}+\alpha_{f},
\end{equation}
where $\alpha_{f}$ is a smearing parameter analogous to $\alpha$.
For $\alpha_{f}=0$ and we obtain in one dimension
\begin{equation}
G^{f}_{d=1}(k)\vert_{\alpha_{f}=0} = \frac{1}{m} -
\frac{2}{m^{2}} \Big[ {\rm ctgh} \frac{m}{2} - i {\rm ctg}
\frac{k}{2} \Big]^{-1} \ , \quad k \in \ ]-\pi ,\pi ].
\end{equation}
This propagator has a zero -- indicating non-locality --
at $k=\pi$, iff $m=0$.

The corresponding 1d perfect scalar propagator can be written as
\begin{equation}
G_{d=1}(k)\vert_{\alpha =0} = \frac{1}{2m} \Big[ 
G^{f}_{d=1}(k)\vert_{\alpha_{f}=0}
+ G^{f}_{d=1}(-k)\vert_{\alpha_{f}=0} \Big] .
\end{equation}
Here we don't find any zeros; in particular 
\begin{equation}
^{\lim}_{m \to 0} G_{d=1}(\pi )\vert_{\alpha =0} = \frac{1}{12}.
\end{equation}
This different behavior of the $\delta $ function FPA
for fermions and scalars persists in higher dimensions.

\begin{figure}[hbt]
\begin{center}
\hspace{7mm}
\def\fpsangle{0}
\epsfxsize=65mm
\fpsbox{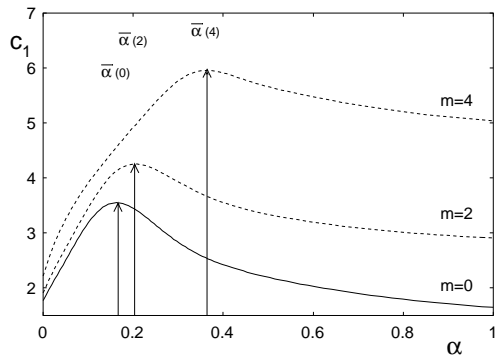}
\end{center}
\vspace{-5mm}
\caption{\it{The decay of the couplings 
$\vert \rho (i,0,0,0)\vert \propto \exp \{ - c_{1} (\alpha ) i\}$ 
of the perfect scalars for masses 0, 2 and 4.
The maxima are very close to the values of $\bar \alpha (m)$, 
0.167, 0.203 resp.\ 0.364, which correspond to eq.\ (\ref{smear}).}}
\vspace{-2mm}
\label{decalp}
\end{figure}

In coordinate space we write the perfect scalar action as
\begin{equation}
S[\phi ] = \frac{1}{2} \sum_{x,y \in \Z^{d}} \phi_{x} 
\rho (x-y) \phi_{y} ,
\end{equation}
where $\rho (r)$ is the inverse propagator, i.e.\ the (inverse) Fourier
transform of $G(k)^{-1}$. Fig.\ \ref{decalp} shows how the exponential
decay coefficient $c_{1}$ on an axis depends on
the smearing parameter $\alpha$, 
$\rho (i,0,0,0) \propto \exp (- c_{1} (\alpha )i)$, at $m=0,\ m=2$ 
and $m=4$. 
\footnote{The curves in Fig.\ \ref{decalp} are Bezier interpolations, 
i.e.\ smooth curves which do not pass through every single data point.
Since the exact curve is not that smooth, the Bezier interpolation 
is appropriate to identify the region of optimal locality.}

\begin{figure}[hbt]
\begin{center}
\hspace{5mm}
\def\fpsangle{270}
\epsfxsize=65mm
\fpsbox{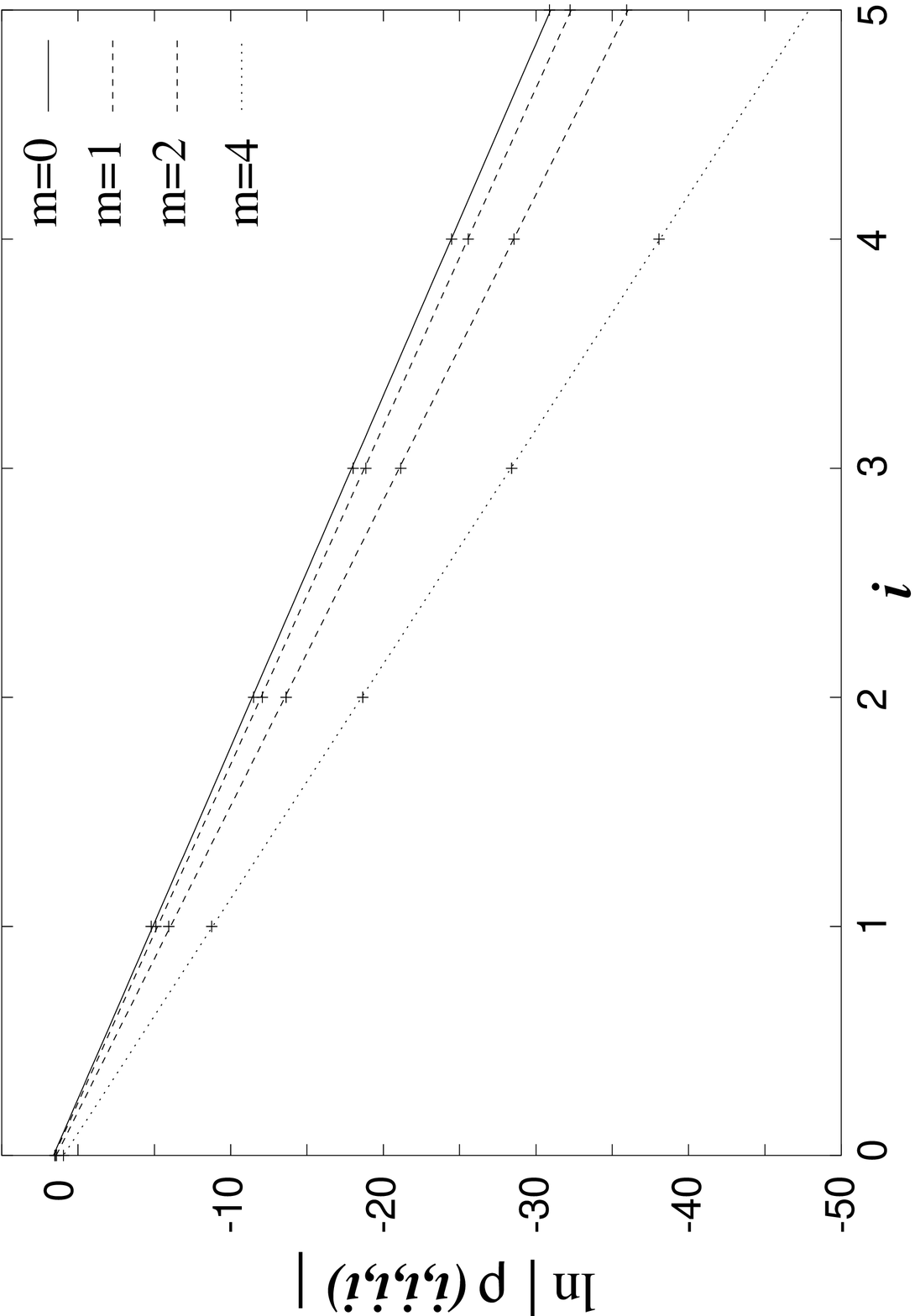}
\vspace{-5mm}
\end{center}
\caption{\it{The decay of the couplings
$\vert \rho (i,i,i,i)\vert $ in the optimally local perfect scalars
for masses 0, 1, 2 and 4.
The lines are least square fits to $\vert \rho (i,i,i,i) \vert
\propto \exp \{ -c_{4}(m)i \}$, with $c_{4}(0) = 6.515, \ c_{4}(1) = 6.775, \
c_{4}(2) = 7.493$ and $c_{4}(4) = 9.769$.}}
\vspace{-2mm}
\label{decexp}
\end{figure}

\begin{figure}[hbt]
\begin{center}
\hspace{15mm}
\def\fpsangle{0}
\epsfxsize=65mm
\fpsbox{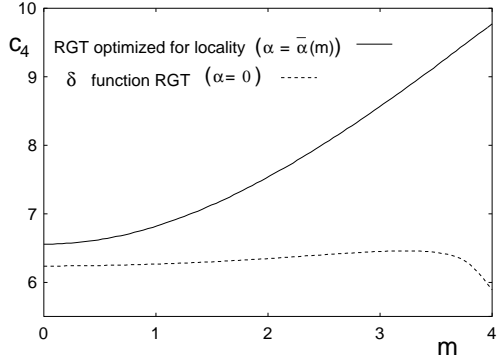}
\vspace{-5mm}
\end{center}
\caption{\it{The decay of the couplings of the perfect scalars
-- measured by $c_{4}$ in 
$\vert \rho (i,i,i,i) \vert \propto \exp \{ - c_{4}i \}$ --
as a function of the mass $m$. The optimally local perfect
scalars follow approximately a parabola, and its decay
is clearly faster the one obtained from a $\delta$ function RGT.}}
\vspace{-2mm}
\label{decmass}
\end{figure}

The figure confirms that $\bar \alpha$ given in eq.\ (\ref{smear})
is an excellent candidate for optimal locality. 
It also affirms that the maximum moves to a larger $\alpha$
as the mass increases.
We stay with the neat analytical criterion for the selection of
$\alpha$ presented above, and focus on the mass dependent 
value of $\bar \alpha$ given in eq.\ (\ref{smear}). Fig.\ \ref{decexp} 
illustrates the exponential decay at various masses. The decay 
becomes even faster if the mass increases, as it was observed 
for fermions before \cite{QuaGlu}. 
This is illustrated in Fig.\ \ref{decmass}, where we consider
the exponential decay $\vert \rho (i,i,i,i) \vert \propto 
\exp (-c_{4}(m)i)$. Especially in the interval $m=0\dots 2$,
the decay coefficients follow very closely a parabola,
\begin{equation}
c_{4}(m) \simeq 6.563 + 0.247 \ m^{2} .
\end{equation}

The largest couplings in 4d coordinate space
for $m=0,\ 1,\ 2$ and 4 are given in Table \ref{copinf}.

\begin{table}
\begin{center}
\begin{tabular}{|c|c|c|c|c|}
\hline
$ r $ & $m=0$ & $m=1$ & $m=2$ & $m=4$ \\
\hline
\hline
(0000) & ~4.54276730 & ~4.46018784 & ~4.11732967 & ~2.58349457 \\
\hline 
(1000) & -0.25656026 & -0.21033619 & -0.12051159 & -0.01815118 \\
\hline
(1100) & -0.06722788 & -0.05285547 & -0.02707061 & -0.00294191 \\
\hline
(1110) & -0.02211001 & -0.01677521 & -0.00778413 & -0.00061988 \\
\hline
(1111) & -0.00828931 & -0.00611485 & -0.00261666 & -0.00015765 \\
\hline
(2000) & -0.00079938 & -0.00042262 & -0.00002726 & ~0.00001090 \\
\hline
(2100) & -0.00058401 & -0.00036971 & -0.00010471 & -0.00000201 \\
\hline
(2110) & -0.00027747 & -0.00018443 & -0.00006107 & -0.00000192 \\
\hline
(2111) & -0.00009663 & -0.00006885 & -0.00002660 & -0.00000098 \\
\hline
(2200) & ~0.00022983 & ~0.00016003 & ~0.00005719 & ~0.00000171 \\
\hline
(2210) & ~0.00013280 & ~0.00008745 & ~0.00002724 & ~0.00000058 \\
\hline 
(2211) & ~0.00008482 & ~0.00005317 & ~0.00001456 & ~0.00000022 \\
\hline
(2220) & ~0.00005018 & ~0.00003084 & ~0.00000804 & ~0.00000010 \\
\hline
(2221) & ~0.00003271 & ~0.00001947 & ~0.00000464 & ~0.00000005 \\
\hline
(2222) & ~0.00001011 & ~0.00000573 & ~0.00000120 & ~0.00000001 \\
\hline
(3000) & -0.00005999 & -0.00003548 & -0.00000842 & -0.00000009 \\
\hline
(3100) & -0.00002060 & -0.00001177 & -0.00000252 & -0.00000002 \\
\hline
\end{tabular}
\end{center}
\caption{\it{The largest couplings $\rho (r)$ of the perfect scalar action
in coordinate space for masses $m=0$, 1, 2 and 4. The table
includes all couplings with absolute values $\geq 10^{-5}$.}}
\label{copinf}
\end{table}

\section{Truncation by periodic boundary conditions}

So far we have worked in an infinite volume, now we consider
a hypercubic finite volume with periodic boundary conditions.
Of course in a finite volume there are no fixed points in the
proper sense, but there are still perfect actions, similar to
the perfect actions at finite mass. 

Let us fix a volume $V$ and $m=0$. Then we start from a
much larger volume, which shrinks under a number of
RGTs (block factor $< \infty$)
just to the fixed final volume $V$. If we now
expand the initial volume more and more, such that the number 
of RGTs gets larger and larger, then the couplings in
$V$ may converge. Their limit represents the perfect action
in the finite volume. (Analogously we can fix a finite ultimate
mass in $V=\infty$ and start from a smaller and smaller initial mass, 
to obtain
after a number of RGTs couplings that converge to those identified 
for the final mass in Section 2.)

The transition from infinite to finite volume is technically
easy: we just replace the integrals over the Brillouin
zone in all the formulae of Section 2 by discrete sums, and
the rest remains unaltered, including the smearing parameter
for optimal locality. We may call this a ``decimation in momentum 
space''.  Of course, the couplings $\rho (r)$ will change a little
compared to $V=\infty$;
for instance an exponential decay is not possible any more.
At the boundary of $V$ they have to vanish smoothly, which suggests
the use of this mechanism as a truncation scheme:
we impose periodic boundary conditions in some volume,
set the couplings to points outside this volume to zero,
and finally use this set of couplings in any volume.

This scheme has a number of virtues: the normalization
conditions for $\sum_{r} \rho (r)$ and $ \sum_{r} r^{2}
\rho (r)$ are precisely fulfilled
(in particular for $m=0$ these sums are 0 resp. $-2d$),
and the transitions
to lower dimensions by summing over the lattice sites
in the supplementary directions is also exact.
Hence the selection criterion for the RGT parameters,
which refers to the mapping on $d=1$, is still sensible
for the truncated system.

All these properties are not obeyed if we truncate 
by just chopping off the couplings in infinite
coordinate space outside a given volume. If we start in this way, we
have to re-adjust the normalizations a posteriori, which 
is 
quite ambiguous.
Moreover, the periodic truncation has a huge practical 
advantage: certain quantities in perfect lattice perturbation theory
take a rather complicated form, such as the perfect quark-gluon vertex
in QCD \cite{QuaGlu,StL}. It is a difficult numerical task to
evaluate them and to transform them to coordinate space, where
they are ultimately needed. If only a few discrete momenta
are involved, this task simplifies tremendously.

\begin{figure}[hbt]
\begin{tabular}{cc}
\def\fpsangle{270}
\epsfxsize=50mm
\fpsbox{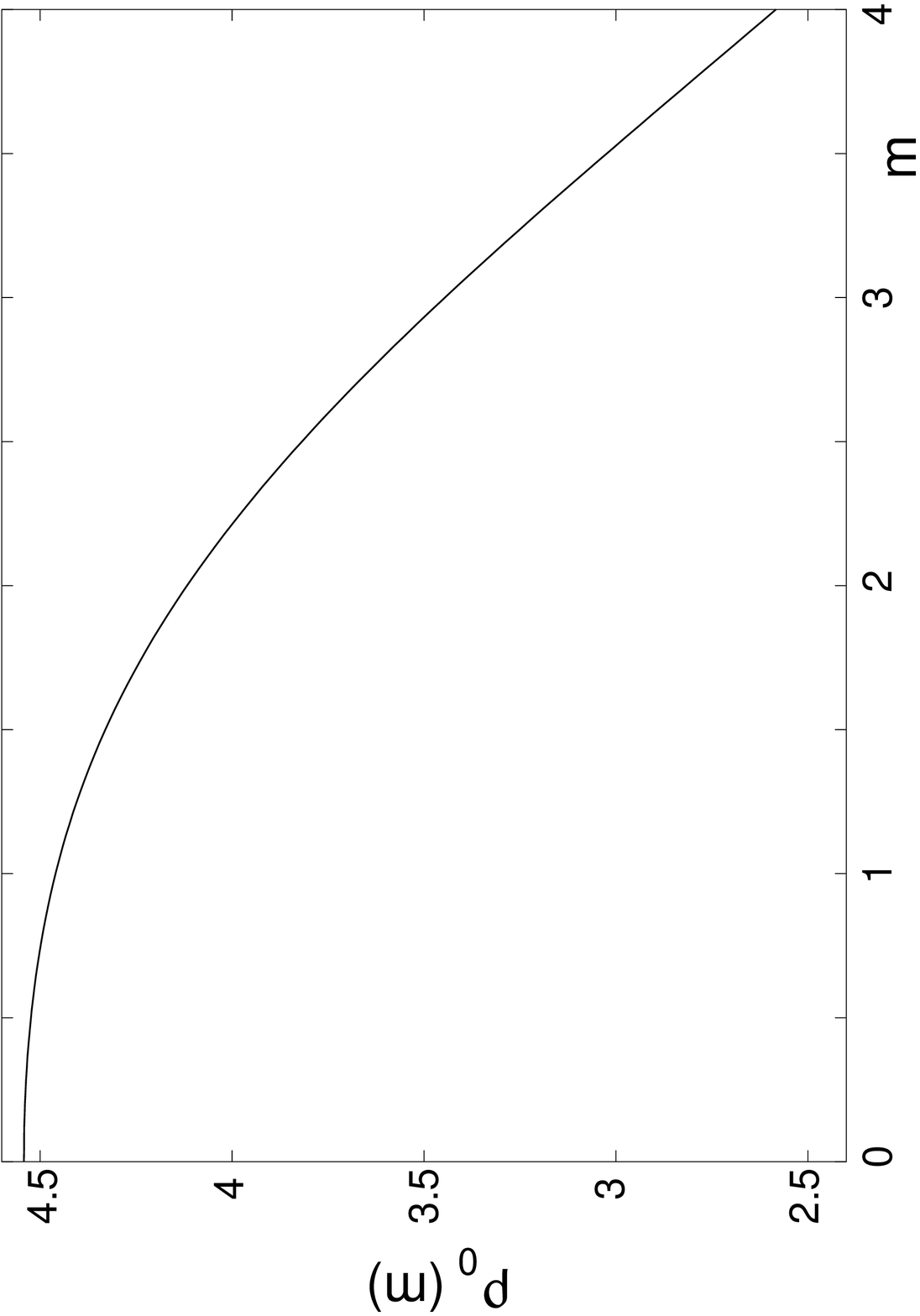}
\hspace{5mm}
\def\fpsangle{270}
\epsfxsize=50mm
\fpsbox{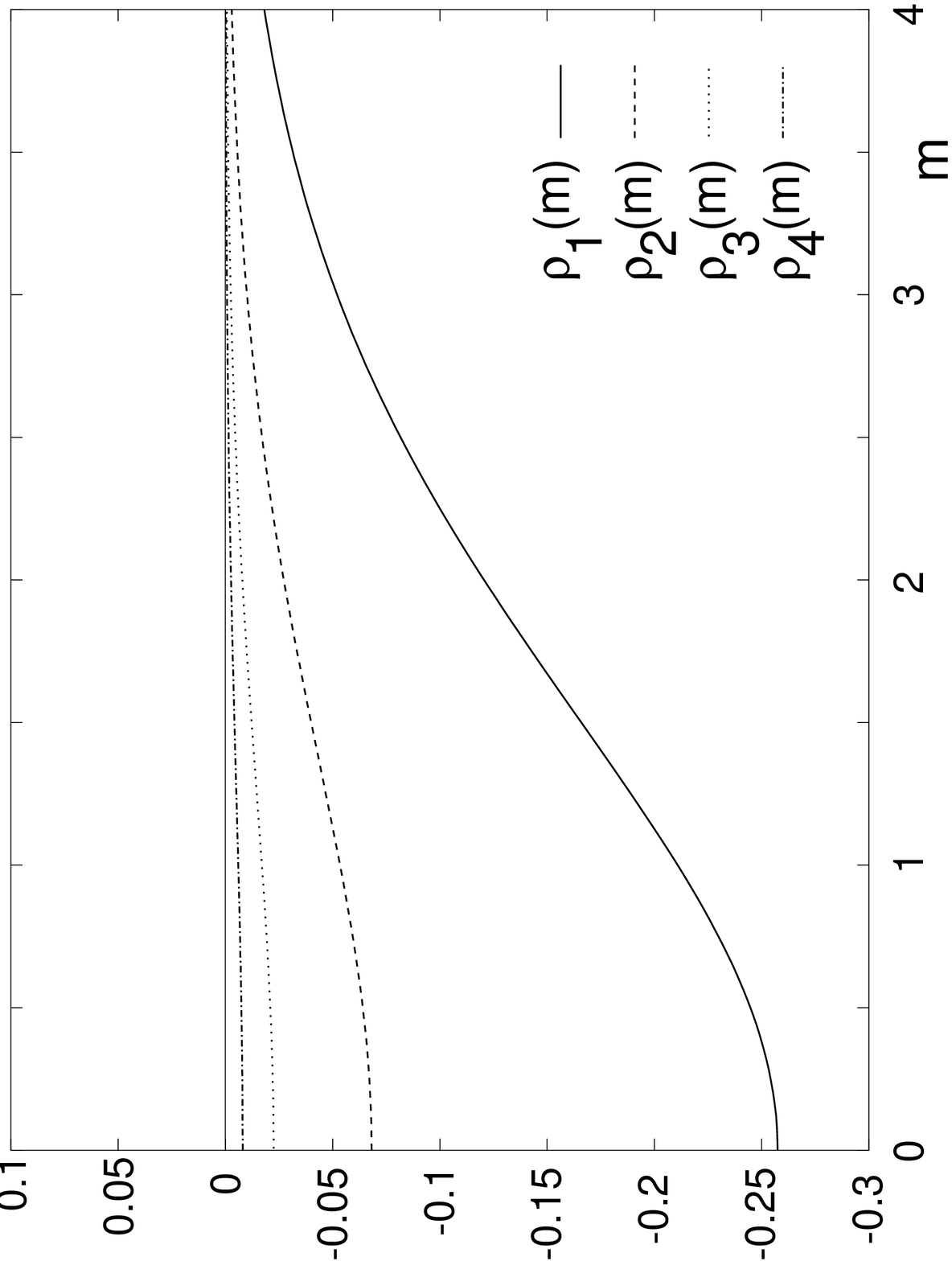}
\end{tabular}
\caption{\it{The mass dependence of the hypercubic couplings
$\rho_{0} = \rho (0,0,0,0)$ (left), as well as
$\rho_{1} = \rho (1,0,0,0)$, $\rho_{2} = \rho(1,1,0,0)$,
$\rho_{3} = \rho (1,1,1,0)$ and $\rho_{4}= \rho (1,1,1,1)$ 
(right).}} 
\vspace{-2mm}
\label{fhypm}
\end{figure}

We apply the preferred procedure by imposing periodic
boundary conditions over a distance of 3 lattice spacings.
Thus the remaining couplings are restricted to a unit
hypercube. There are formally $3^{d}$ of them, but they come
in just $d+1$ equivalence classes. Furthermore the mapping
condition on the ultralocal 1d case and $\sum_{r} r^{2} \rho (r)$
impose three constraints
on the equivalence classes, such that e.g.\ in $d=4$ there
remain only two degrees of freedom. We give the full set of
$d+1$ couplings in $d=2, \ 3$ and 4 for masses $m=0,\ 1,\ 2$
and 4 in Table \ref{copvol}, and we illustrate the smooth
mass dependence of the truncated 4d couplings in Fig.\ \ref{fhypm}.
This truncation range is of practical interest, since it has
been shown that hypercube particles are tractable in 4d 
simulations \cite{Org,HF,precon}.

\begin{figure}[hbt]
\begin{tabular}{cc}
\hspace{-2mm}
\def\fpsangle{270}
\epsfxsize=50mm
\fpsbox{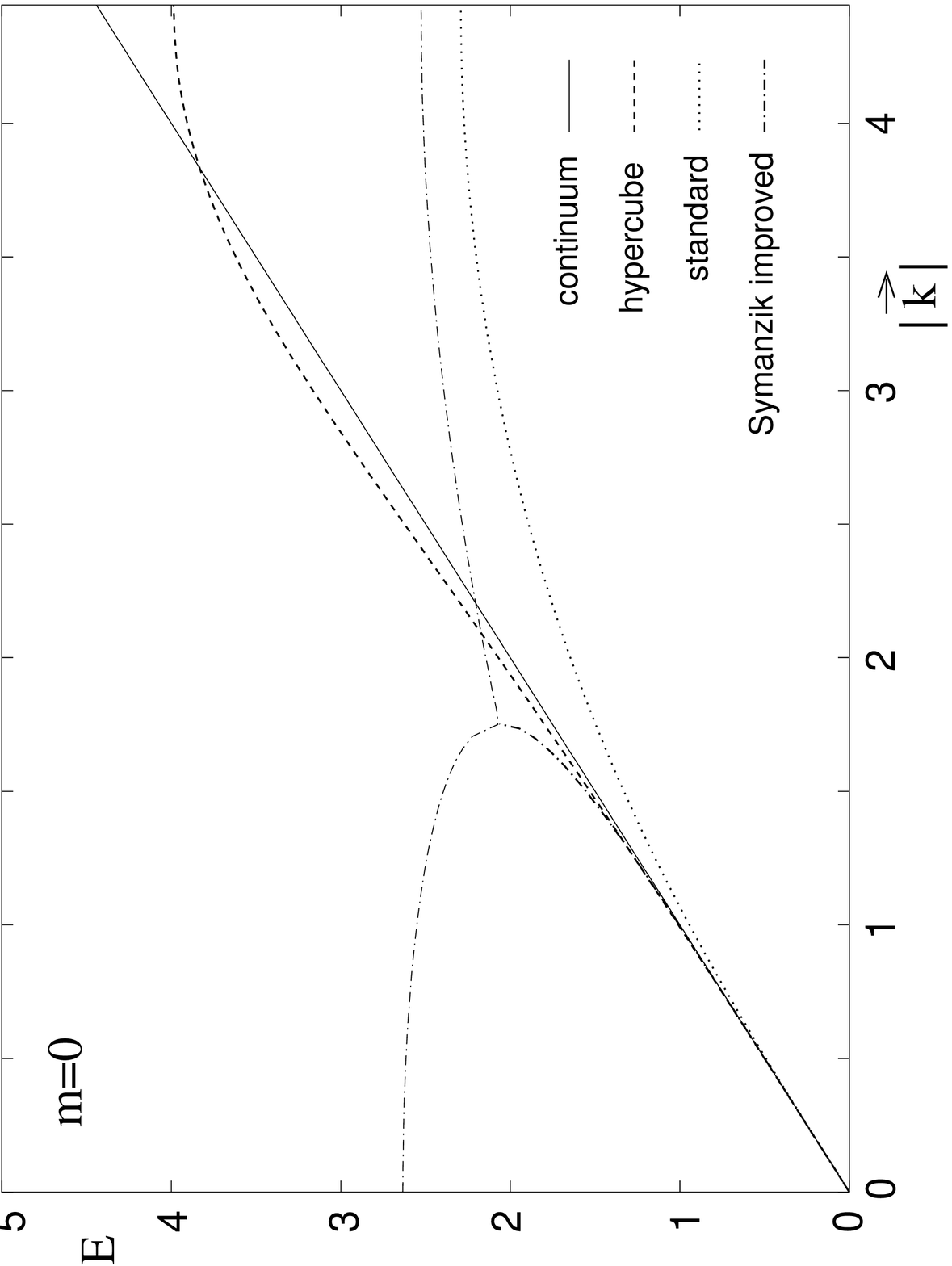}
\def\fpsangle{270}
\epsfxsize=50mm
\fpsbox{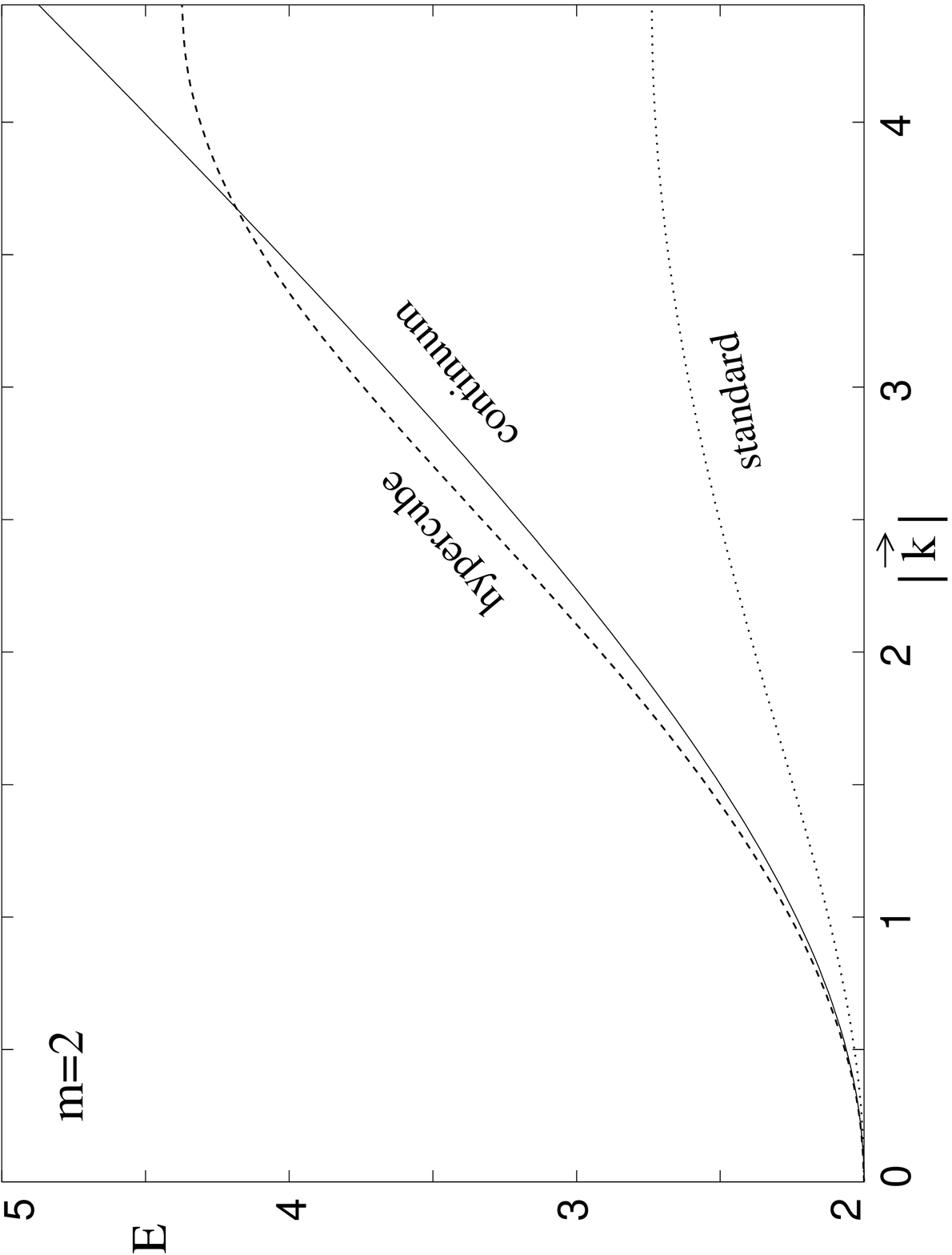}
\end{tabular}
\caption{\it{The dispersion relation for a scalar of mass 
$m=0$ (left) and $m=2$ (right) in the (110) direction
with the perfect, truncated perfect (``hypercubic'')
and standard lattice action.
For the hypercube action (as well as the standard action) there
are no ``ghosts'' (higher branches) in any direction, in contrast
to the Symanzik improved action.}}
\vspace{-2mm}
\label{fdisp}
\end{figure}

One may argue that the really perfect actions of Section
2 (in $d>1$) are academic, since we always need
some truncation for practical purposes.
As a first check on how much the perfect properties suffer
from the 3-periodic truncation, we show the dispersion
relation of the ``hypercube scalar'' for $m=0$ and $m=2$ in
Fig.\ \ref{fdisp}, where they are compared to the perfect or
continuum spectrum and to the spectrum obtained from
the standard lattice action. 
As an example, we consider the direction (110) and we observe a dramatic
improvement, all the way up to the edge of the Brillouin zone.
The situation is similar in the (100) and the (111) direction.
This is a success of the optimized locality in Section 2.
Note that the truncated perfect energies start with a slight
overshoot at small momenta. Following Symanzik's improvement 
program -- which is fundamentally different from the program
discussed here -- it would be possible to correct the leading order
artifacts $\propto \vec k ^{2}$. Then the dispersion becomes even better
for small momenta, but with respect to larger momenta
this is not necessarily an advantage. At the edge of the Brillouin
zone the curve has to bend down due to periodicity,
so it can be favorable to start with a slight 
overshoot, in order to follow the perfect curve over a wide range.
After all, if one wants to simulate on very coarse lattices,
then not only the momenta $\vert \vec k \vert \ll 1$ are important.
We also emphasize that our ``hypercube scalar'' follows directly
from a very general prescription; no particular tuning
for a good dispersion is involved. This raises hope that the 
same prescription is also successful for other quantities.\\

For comparison, we consider a Symanzik improved scalar by
including additional couplings along the axes over two
lattice spacings. The action
\begin{equation}
S[\phi ] = \frac{1}{2} \sum_{x,y} \phi_{x} \Big[
\delta_{x,y} \Big( \frac{8d}{3} + m^{2} \Big) 
+ \sum_{\mu} \Big( -\frac{4}{3} \delta_{x+\hat \mu ,y}
+ \frac{1}{12} \delta_{x+2\hat \mu ,y} \Big) \Big] \phi_{y}
\end{equation}
is $O(a^{2})$ improved; the remaining artifacts for the
free scalars are of $O(a^{4})$.
The dispersion relation of this Symanzik 
improved scalar is also shown in Fig.\ \ref{fdisp} (left).
By construction it is close to the continuum scalar for
small momenta, but at $k_{1}=k_{2}=1.239$ it is hit by
a higher branch, and the continuation of the curve is just the real 
part of two complex conjugate solutions, which are useless.

In exactly the same way one can improve the staggered fermions
and arrives at the Naik fermion \cite{Naik}, which combines
the discrete derivatives to nearest neighbors and over distances
of three lattice spacings with the relative weights 9/8 and $-1/24$.
A similar construction starting from the Wilson fermion
is called D234 action \cite{D234}.
It turns out that its behavior is very similar to the one
of the Symanzik improved scalar shown in Fig.\ \ref{fdisp} (left).
Also Naik fermions and D234 fermions are
hit by higher branches at $\vert \vec k \vert =O(1)$, which marks 
the end of the real solution for $E(\vec k )$. Those dispersions
are compared to the truncated perfect fermions in Ref.\ \cite{stag,StL}.

Similarly, if we truncate an optimally local,  perfect free gauge 
field to a 4d hypercube, we find an excellent (transverse)
dispersion relation \cite{StL}.

\begin{table}
\begin{center}
\begin{tabular}{|c|c|c|c|c|}
\hline
$ r $ & $m=0$ & $m=1$ & $m=2$ & $m=4$ \\
\hline
\hline
$d=2$ & & & & \\
\hline
(00) & ~3.23975790 & ~3.40289300 & ~3.52616893 & ~2.49915565 \\
\hline 
(10) & -0.61987895 & -0.49256958 & -0.26092482 & -0.03241219 \\
\hline
(11) & -0.19006053 & -0.14542414 & -0.06917628 & -0.00607959 \\
\hline
$d=3$ & & & & \\
\hline
(000) & ~4.02729212 & ~4.03823126 & ~3.87615269 & ~2.54721313 \\
\hline 
(100) & -0.39376711 & -0.31766913 & -0.17499188 & -0.02402874 \\
\hline
(110) & -0.11305592 & -0.08745023 & -0.04296647 & -0.00419173 \\
\hline
(111) & -0.03850230 & -0.02898696 & -0.01310490 & -0.00094393 \\
\hline
$d=4$ & & & & \\
\hline
(0000) & ~4.54224606 & ~4.45988071 & ~4.11725747 & ~2.58349386 \\
\hline 
(1000) & -0.25747697 & -0.21082472 & -0.12055239 & -0.01814036 \\
\hline
(1100) & -0.06814507 & -0.05342220 & -0.02721974 & -0.00294419 \\
\hline
(1110) & -0.02245542 & -0.01701401 & -0.00787336 & -0.00062377 \\
\hline
(1111) & -0.00802344 & -0.00598647 & -0.00261577 & -0.00016008 \\
\hline
\end{tabular}
\end{center}
\caption{\it{The couplings $\rho (r)$ of the truncated perfect 
scalar action in coordinate space for masses 
$m=0$, 1, 2 and 4.}}
\label{copvol}
\end{table}

\section{Thermodynamic properties}

In thermodynamics the artifacts due finite lattice spacing are
particularly bad \cite{FN}, hence the use of improved actions is
strongly motivated. Thermodynamics with quasi-perfect
lattice actions has also been studied for pure $SU(3)$ gauge 
theory \cite{papa}, and for the 2d $O(3)$ model \cite{spiegel}.

First we consider massless scalars with $N_{t}$ lattice points
in the Euclidean time direction. We impose periodic boundary
conditions in the 4 direction over those $N_{t}$ lattice spacings.
The pressure $p$ on an infinite lattice with the standard action 
is given by \cite{bielalt}
\begin{eqnarray}
\frac{p}{T^{4}} &=& \frac{N_{t}^{4}}{(2\pi )^{3}} \int_{-\pi}^{\pi} d^{3} k 
\ \Big[ \frac{1}{N_{t}} \sum_{n=1}^{N_{t}} \ln \Delta(\vec k ,
k_{4,n})^{-1}\vert_{k_{4,n} =2\pi n/N_{t}} \nonumber \\ 
&& - \frac{1}{2\pi }\int_{-\pi}^{\pi} d k_{4} \ 
\ln \Delta(\vec k, k_{4})^{-1} \Big] , \label{pres}
\end{eqnarray}
where $T$ is the temperature.

\begin{figure}[hbt]
\begin{tabular}{cc}
\def\fpsangle{0}
\epsfxsize=70mm
\fpsbox{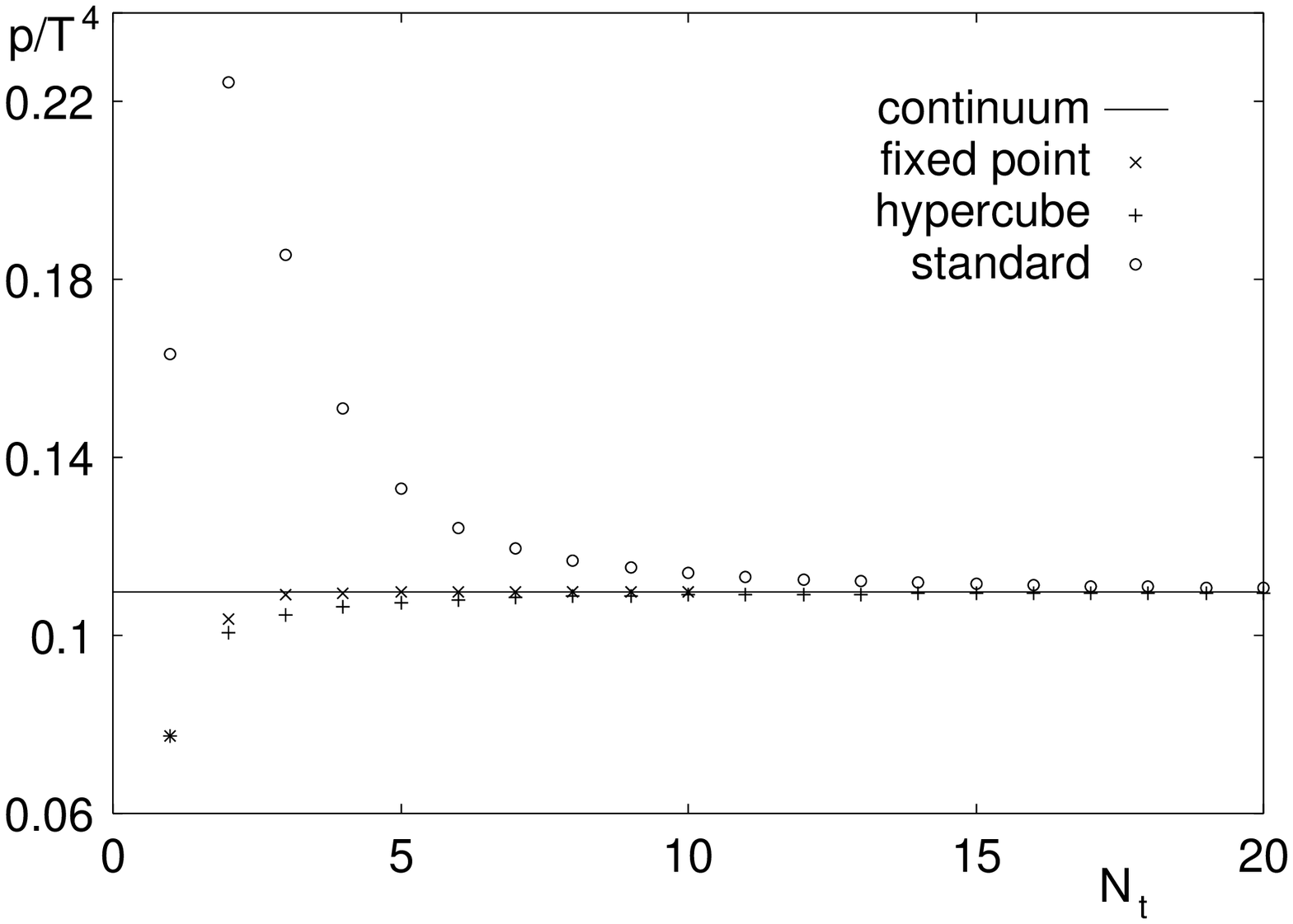}
\def\fpsangle{270}
\epsfxsize=50mm
\fpsbox{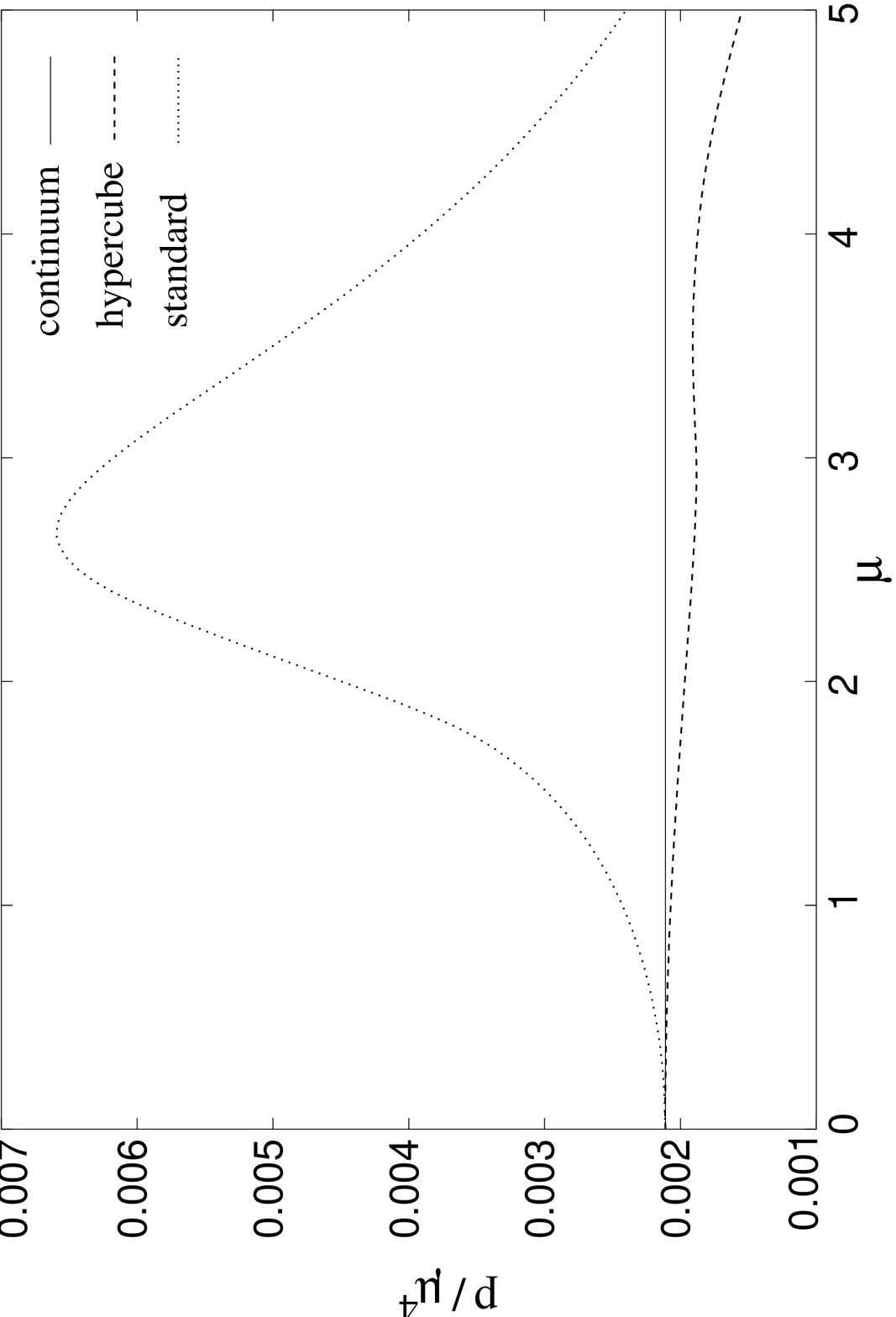}
\end{tabular}
\vspace{-5mm}
\caption{\it{Left: The ratio $p/T^{4}$ at $N_{t}$ discrete points
in the Euclidean time direction for
the perfect, the truncated perfect and the standard lattice 
action. 
Right: the scaling ratio $p/\mu^{4}$ ($\mu$:
chemical potential) at $T=0$ in the continuum, compared to
different lattice actions.}}
\vspace{-2mm}
\label{fpres}
\end{figure}

To study the improvement of the lattice actions from Section 2 and
3, we just replace the standard lattice propagator in 
eq.\ (\ref{pres}) by the the fixed point or truncated
fixed point propagator. 
In Fig.\ \ref{fpres} the results for the three lattice actions are
compared to the Stefan Boltzmann law in the continuum,
\begin{equation}
\frac{p}{T^{4}} = \frac{\pi^{2}}{90} .
\end{equation}
In this respect even the ``perfect actions''
contain lattice artifacts. The reason is that we ignored
``constant factors'' when we performed the Gaussian
functional integrals over $\D \tilde x$ and $\D \tilde \sigma$
in Section 2.
For instance in the spectrum such factors have indeed no
influence. However, they can depend on the temperature,
yielding artifacts in thermodynamics. These artifacts 
are exponentially suppressed, and -- as the Figures show --
they disappear very fast as $N_{t}$ increases. The truncation
slows down a little the convergence to the Stefan Boltzmann value,
but the behavior is still strongly
improved compared to the standard lattice action.
The latter suffers from quadratic artifacts (asymptotically
$0.38 \pi^{2}/N_{t}^{2}$). Hypercubic
truncation also yields quadratic artifacts, but only of about
$-0.05 \pi^{2}/N_{t}^{2}$. \\

As a second example, we return to $T=0$ but introduce a 
chemical potential $\mu$ according to the instruction for
general lattice actions given in Ref.\ \cite{chem}.
We then measure the scaling ratio $p/\mu^{4}$ for massless
scalars. In the continuum
it amounts to $1/48 \pi^{2}$, and all lattice actions approach
this value in the limit $\mu \to 0$. For finite $\mu$ the lattice 
artifacts are visible. Fig.\ \ref{fpres} (right) 
shows that they are dramatic for the standard action, 
whereas the artifacts in the hypercube action are again quite modest.

\section{Preconditioning the hypercube scalar}

If we want to simulate a hypercube scalar action, 
we are interested in efficient ways to compute the
scalar determinant, analogous to the fermion determinant.
There are a number of well established algorithms for this
purpose, for an overview see Ref.\ \cite{frommer}. The
maximal speed of such methods is limited by the eigenvalue 
distribution. The goal of ``preconditioning'' is to transform
the matrix such that its spectrum becomes more favorable with 
this respect. In particular, it is of advantage to squeeze 
the eigenvalues into a narrow interval.

In the case of the standard action we can order the matrix lines
by passing successively through the even and the odd
sub-lattice, and we obtain a matrix of the block form
\footnote{We re-scale the matrix such that the diagonal elements 
become 1.}
\begin{equation}
M = \left( \begin{array}{cc} \unit & -A \\ -A & \unit \end{array} \right) .
\end{equation}
In a volume $V$, $M$ is a $V \times V$ matrix
which splits into 4 blocks of the same size ($\unit$ represents
unity). We decompose $M$ into $M=\unit -L-U$, where $L$, $U$
are the strictly lower resp.\ upper triangular part of $M$. 
Now we define the matrices $V_{1} := \unit -L$,
$V_{2} := \unit -U$, which can be inverted trivially.
The SSOR and the ILU preconditioning amount
in this case both to the transformation
\begin{equation}
M' = V_{1}^{-1} M V_{2}^{-1} = \left( \begin{array}{cc}
\unit & 0 \\ 0 & \unit - A^{2} \end{array} \right) \ , \label{decom}
\end{equation}
where ${\rm det} M' = {\rm det} M$. To solve a linear system of equations
given by $M$ we can now apply the Eisenstat trick \cite{eisen},
where we only need to invert triangular matrices. Moreover,
referring to our criterion mentioned above, we can clearly expect
the spectrum of $M'$ to be closer to 1 than the spectrum of $M$.

For the hypercube scalar the situation is more complicated. 
To obtain a block structure in $M$, we have to
decompose the lattice into $2^{d}$ sub-lattices, which we
go through successively. This sub-lattice structure is known from
staggered fermions. 
We denote this extension of the ``red-black'' scheme as
``{\em rainbow preconditioning}\,''.
Thus we obtain $2^{d}$ unit blocks along the
diagonal of $M$, and many more nonzero elements than it
is the case for the standard action. They arise from the $d$
hopping parameters. 
We denote the maximal magnitude of the elements in $L$ and $U$
as $O(\varepsilon )$.
We still apply the multiplication (\ref{decom}) and arrive at
\begin{equation}\label{decohyp}
M' =\unit - ( \sum_{i\geq 1}^{V} L^{i} ) \ ( \sum_{j\geq 1}^{V}
U^{j}) = \unit - LU - O(\varepsilon^{3}) .
\end{equation}
We cannot eliminate the off-diagonal elements to such a large
extent as it was possible for the standard action, because there
are more of them from the beginning. On the other hand,
we can still expect the spectrum of $M'$ to be much closer to
1, typically the eigenvalues are $1-O(\varepsilon^{2})$. Since
$\varepsilon$ is clearly suppressed for the hypercube scalar --
compared to the $\varepsilon$ occurring in
the standard action -- the suppression of the
deviation from the unit matrix ($O(\varepsilon )$ for $M$, 
$O(\varepsilon^{2})$ for $M'$) becomes even more powerful.
In addition we are still in 
agreement with the conditions for the Eisenstat trick.

\begin{figure}[hbt]
\begin{tabular}{cc}
\def\fpsangle{270}
\epsfxsize=50mm
\fpsbox{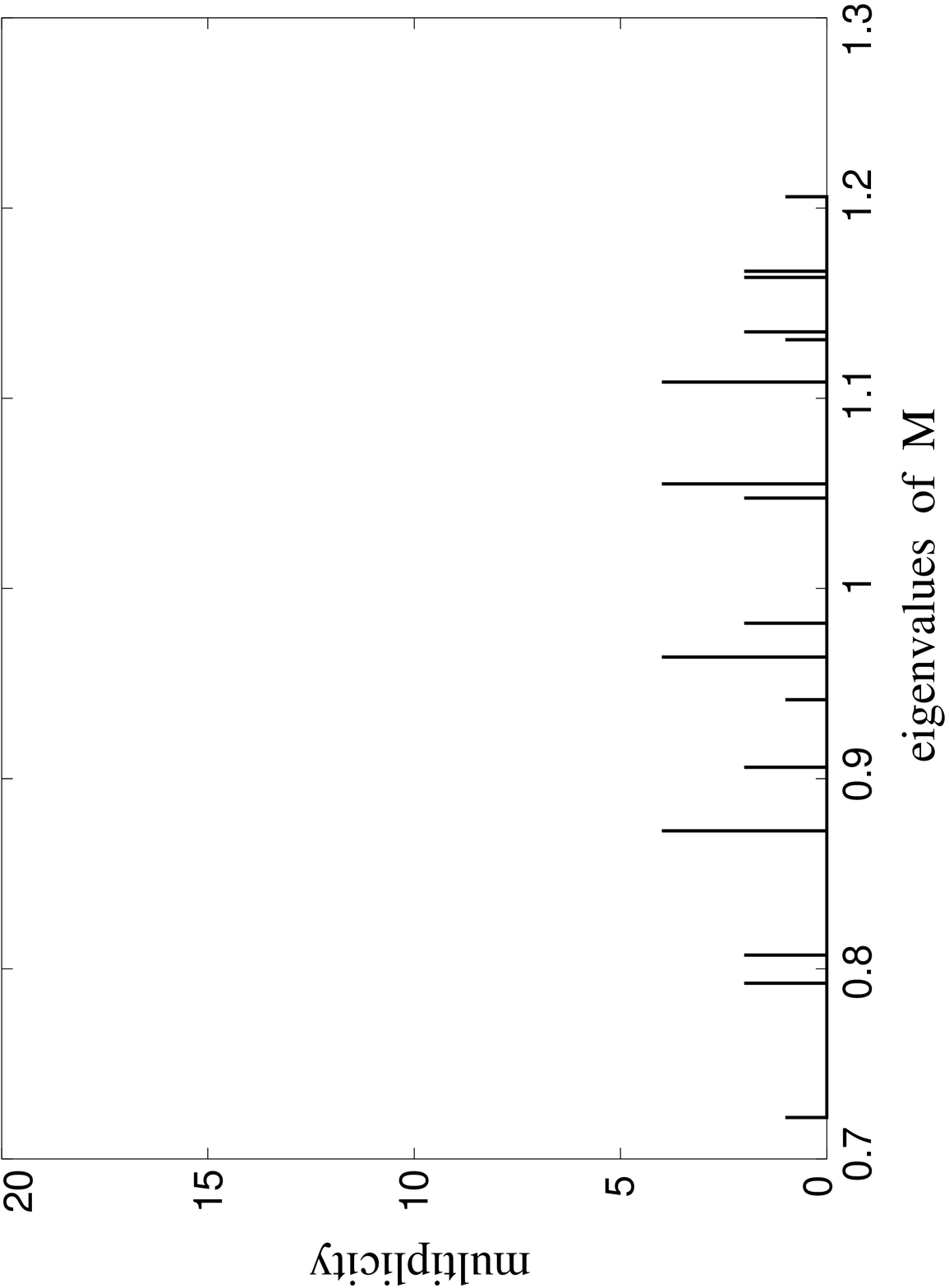}
\def\fpsangle{270}
\epsfxsize=50mm
\fpsbox{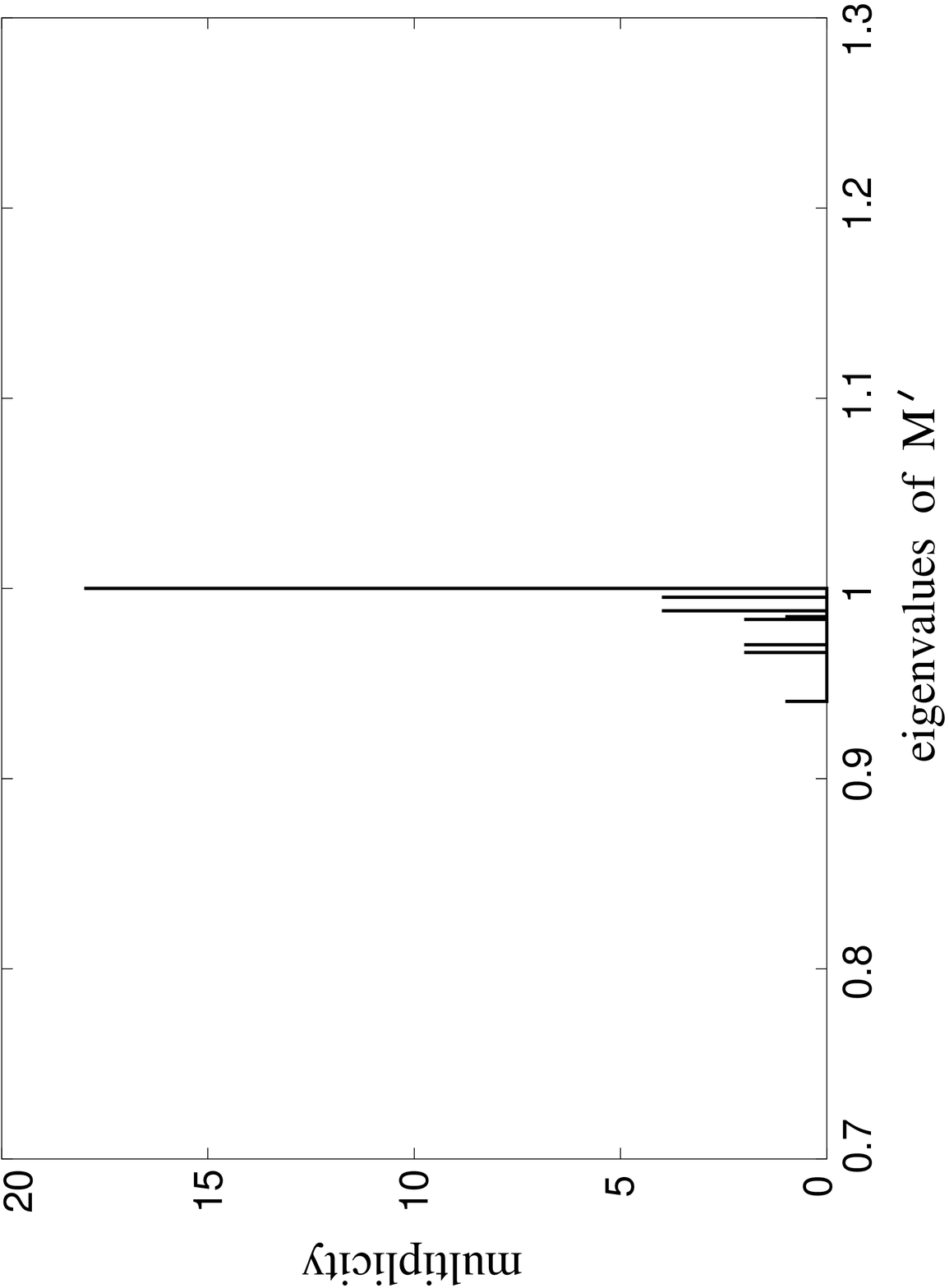}
\end{tabular}
\vspace{-3mm}
\caption{\it{The eigenvalue distribution of the hypercube scalar matrix $M$
-- before preconditioning -- (left),
and of the corresponding matrix $M'$ -- after preconditioning -- (right)
on a $6 \times 6$ lattice.}}
\label{specfig}
\end{figure}

To observe this effect, we consider the 2d case for $m=1$, where
the off-diagonal elements are 0, $-0.072375$ or $-0.0213675$, as we 
know from Table \ref{copvol}.
On a $6\times 6$ lattice one obtains the spectrum illustrate in
Fig.\ \ref{specfig} for $M$ (left), and for $M'$ (right).
The hight of the lines represents the algebraic multiplicity
of the corresponding eigenvalues. We see that the eigenvalues of
$M'$ are indeed concentrated close to 1, and all of them are
$\leq 1$, as one may suspect from eq.\ (\ref{decohyp}). In particular,
the eigenvalue most distant from 1 is 0.72185 for $M$, and it
moves to $0.94060$ for $M'$. Note also that $M$ does not
have an eigenvalue 1, but in $M'$ there are 18 of them
(plus 2 at 0.99537, which are hidden in Fig.\ \ref{specfig}).
The ``squeezing'' effect is indeed stronger for the truncated 
perfect action than for the standard action.

We conclude that preconditioning is still applicable and efficient
for hypercube actions. The same holds for hypercube fermions
(perfect fermions truncated to a unit hypercube);
for first experiments in QCD, see Ref.\ \cite{precon}.

\section{Non hypercubic lattices}

\subsection{Anisotropic lattices}

It is straightforward to write down the perfect
free scalar propagator on arbitrary hyper-rectangular lattices.
We call the lattice spacing in $\mu$ direction $a_{\mu}$,
$\mu = 1, \dots d$. Then the blocking from the continuum
involves an integral over the hyper-rectangular lattice cells.
The smearing parameter $\alpha$ is also affected, since it
has dimension $mass^{-2}$.
We have to make a choice, for which 1d projection we want
to obtain ultralocality. If this is the case for the mapping
on the $\nu$ axis, then the generalized perfect propagator reads
\begin{eqnarray}
G_{ani}(k) &=& \sum_{l \in \Z^{d}} \Delta (k+2\pi (l/a)) 
\Pi^{2} (k+2\pi (l/a))
+ \frac{\bar \alpha}{a_{\nu}^{2}} \nonumber \\
(l/a) & := & (l_{1}/a_{1}, \dots , l_{d}/a_{d}) \nonumber \\
\Pi (k) &=& \prod_{\mu =1}^{d} \frac{\hat k_{\mu}}{k_{\mu}} ,
\quad \hat k_{\mu} := \frac{2}{a_{\mu}} \sin 
\frac{k_{\mu}a_{\mu}}{2} ,
\end{eqnarray}
where $k \in B = \ ] -\pi /a_{1},\pi/a_{1}] \times \dots
\times \ ]-\pi /a_{d}, \pi/a_{d}]$, and $\bar \alpha$ is still 
the mass dependent smearing parameter from eq.\ (\ref{smear}).
Note that the Poincar\'{e} invariance of the spectrum
persists, whatever the lattice spacings.\\

Anisotropic lattices are often used in thermodynamics.
To simulate at finite temperature on isotropic lattices,
there are typically just $N_{t}=4$ to 8 points available in the
Euclidean time direction. In order to measure
correlators precisely one would like to have a larger
$N_{t}$. The anisotropy permits to invest a larger
fraction of the available lattice points in the 4 direction
(in a hypercubic volume).
One introduces a spatial lattice constant $a_{\sigma}$,
and a temporal one $a_{t}<a_{\sigma}$.

If one then measures masses in all directions, one recovers
isotropy up to lattice artifacts, which can be used as a test of
improved actions in this context. If the improvement is
successful, one may use a coarse spatial lattice, 
but a small $a_{t}$.
The dispersion relations are the same as on the hypercubic lattice,
up to a rescaling of the axes. Hence the strong improvement
that we observed for the (truncated) perfect scalars persists.

In improved actions, the presence of ``ghosts'' (higher branches in 
the dispersion relation) can cause some trouble because 
they distort the Hermiticity of the transfer matrix. 
In the fully perfect action there are necessarily an infinite
number of such ghosts -- corresponding to the sum over $l$.
In Symanzik improved actions there are usually a few ghosts,
and by some modifications it is possible to get rid of some
of them or all of them, if one takes in account other disadvantages
(for ``D234 fermions'' this has been discussed in Ref.\ \cite{D234}).
The hypercube scalars -- as well as all hyper-rectangular scalars --
have the virtue that they are free of ghosts to start with, 
cf.\ Fig.\ \ref{fdisp}.



\subsection{Triangular lattices}

By blocking from the continuum we can also derive efficiently perfect
actions on triangular lattices or honey comb lattices.
Such actions are completely unexplored so far.
As an illustration we consider the case of a 2d lattice
built from regular triangles with lattice spacing 1. 
Here the blocking from the continuum involves
integrals over the hexagonal lattice cells, which 
form the dual lattice.
\footnote{For the perfect action on a 
honey comb lattice it is the other way round, the
cells to be integrated over are triangles in two different
orientations.}
In coordinates of the suitable axes -- with an angle
of $\pi /3$ -- we find that the $\Pi$ function introduced
in Section 2 is replaced by
\begin{equation}
\Pi_{tria}(k) = \frac{8}{3 k_{1}k_{2}(k_{1}+k_{2})}
\, \Big[ \, k_{1} \cos \frac{k_{1}}{2} + k_{2} \cos \frac{k_{2}}{2}
- (k_{1}+k_{2}) \cos \frac{k_{1}+k_{2}}{2} \, \Big] .
\end{equation}
We recognize the exchange symmetry of the axes and the
special r\^{o}le of the edges of the triangular Brillouin zone.
Like the hypercubic $\Pi$ function, also $\Pi_{tria}$
only contains removable singularities, and in the
limit $k_{1},k_{2} \to 0$ it becomes 1, which confirms
the normalization.

If we only let $k_{2}\to 0$, we arrive at
\begin{equation}
\Pi_{tria}(k_{1},0) = \frac{8}{3 k_{1}^{2}}
\Big[ \, 1 - \cos \frac{k_{1}}{2} + \frac{k_{1}}{2}
\sin \frac{k_{1}}{2} \, \Big] .
\end{equation}
Remarkably, this is different from $\Pi (k_{1})$.\\

The blocking constraint
\begin{equation}
\phi_{x} = \int_{H_{x}} d^{2}y \ \varphi (y)
\end{equation}
-- $H_{x}$ being the hexagon with unit diameter and center $x$ --
takes in momentum space the form
\begin{eqnarray} \label{triacon}
\phi (k) &=& \frac{3}{4} \sum_{l\in \Z^{2}} \varphi (k_{l}) 
\Pi_{tria} (k_{l}) , \nonumber \\
k_{l} & := & k + \frac{4\pi}{3} \left( \begin{array}{cc}
2 & -1 \\ -1 & 2 \end{array} \right) l := k + \frac{4\pi}{3} l'.
\end{eqnarray}
Note that $l'$ does not extend over $\Z^{2}$; it only covers
the points where $l_{1}' - l_{2}'$ is an integer multiple of 3.
These points build again a triangular lattice. They are the centers
of the Brillouin zones, which have a hexagonal form. The lattice 
momentum $k$ in eq.\ (\ref{triacon}) is in the Brillouin zone around
zero, namely the hexagon with the corners $(\pm 4\pi /3 ,0)$,
$(0,\pm 4\pi /3)$, $(4\pi /3,-4\pi /3)$ and $(-4\pi /3,4\pi /3)$.
Thus the integral over $k$, together with the sum over $l'$,
just covers the continuum momentum space. Note that this
structure agrees with that fact that a plane wave along one
axis may have momentum $4\pi$ before being caught in periodicity.\\

Considering that the scalar product is transformed in this
coordinate system, we arrive at the perfect triangular
propagator
\begin{equation}
G_{tria}(k) = \sum_{l\in \Z^{2}} \frac{\Pi_{tria}^{2}
(k_{l})}{k_{l,1}^{2}+k_{l,2}^{2}+k_{l,1}k_{l,2}+m^{2}}
+ \alpha \ .
\end{equation}
The further steps -- evaluation of the inverse propagator in coordinate
space, truncation etc. -- are straightforward, in analogy to Sections
2 to 4.

\section{B-spline blocking}

We return to the hypercubic lattice and we want to explore some
new versions of the blocking from the continuum. The hope is to
find convolution functions, which have advantages over
the simple prescription described in Section 2, particularly
in view of locality.

First we introduce the wavelet or B spline language \cite{wavelet}
in $d=1$, in a dialect which is suitable for our purposes.
The zeroth B spline function is just a $\delta$ function,
\begin{equation}
f_{0}(x) = \delta (x).
\end{equation}
Inductively we proceed from order $N$ to $N+1$ as follows:
\begin{equation}
f_{N+1}(x) = \int_{-\infty}^{x} dy \ \Big[ f_{N}(y+\frac{1}{2})
- f_{N}(y-\frac{1}{2}) \Big] = \int_{x-1/2}^{x+1/2} f_{N}(y) \ dy .
\end{equation}
This implies for instance
\begin{eqnarray}
f_{1}(x) &=& \Big\{ \begin{array}{cc} 1 & \vert x \vert < 
\frac{1}{2} \\ 0 & {\rm otherwise} \end{array} \ , \qquad
f_{2}(x) = \Big\{ \begin{array}{cc} 1- \vert x \vert & 
\vert x \vert < 1 \\ 0 & {\rm otherwise} \end{array} \nonumber \\
f_{3}(x) &=& \left\{ \begin{array}{cc}
\frac{3}{4} -  x^{2} &  \vert x \vert < \frac{1}{2} \\
\frac{1}{2} ( \frac{3}{2} - \vert x \vert )^{2} &
\frac{1}{2} \leq \vert x \vert < \frac{3}{2} \\
0 & {\rm otherwise} \end{array} \right. \ , \qquad {\rm etc.}
\end{eqnarray}
We note a few properties:
$f_{N}(x) = f_{N}(-x)$, $f_{N}(\pm \infty ) = 0$, $f_{N}(x) \geq 0$,
$\int_{-\infty}^{\infty} f_{N}(x) dx = 1$, $f_{N}(x)$
is maximal in $x=0$ and consists of monomial pieces of
order $N$. Furthermore we observe
\begin{equation} \label{demo}
\sum_{j \in \Z} f_{N}(x+j) \equiv 1 \qquad ( {\rm for~all} \ x, \ N \geq 1 ).
\end{equation}
For the periodic blocking from the continuum on a unit lattice,
the means that all continuum points contribute with the same
weight to the lattice variables (``democracy among the continuum 
points'').

For increasing $N$ the functions seem flatter and smoother 
(although $f_{3}$ is still not continuously differentiable).
The support ranges from $-N/2$ to $N/2$.

In $d$ dimensions we simply define $f_{N}(x) = \prod_{\mu =1}^{d}
f_{N}(x_{\mu})$. If we use $f_{N}(x)$ as the convolution function for
blocking matter fields from the continuum, then it is sensible
to insert
\begin{equation}
f_{N,\mu} := f_{N+1}(x_{\mu}) \prod_{\nu \neq \mu} f_{N}(x_{\nu})
\end{equation}
for the non-compact Abelian gauge field $A_{\mu}$. 
This just provides the connection of nearest neighbor lattice 
matter field variables, i.e.\ we integrate over all the straight
continuum connections of corresponding points in adjacent lattice 
cells. This generalizes the prescription applied in Ref.\ \cite{QuaGlu}.\\

The simplest approach in this framework is the decimation
RGT (in coordinate space), which defines the lattice variables as
\begin{equation}
\phi_{x} = \int d^{d}y \ f_{0}(x-y) \varphi (y) \ , \quad
x \in \Z^{d} .
\end{equation}
However, the corresponding expression for the free propagator in
momentum space,
\begin{equation}
G_{0}(k) = \sum_{l \in \Z^{d}} \frac{1}{(k+2\pi l)^{2} + m^{2}}+\alpha
\end{equation}
only converges in $d=1$. 
Still the blocking scheme which arises for
gauge fields in this approach occurs in the literature 
\cite{Het}. On the other hand, for all $N\geq 1$ the convergence of the sum
in the propagator is guaranteed, both, for $f_{N}$ and for $f_{N,\mu}$.\\

What has been done so far in this paper, and in Refs.\
\cite{Schwing,QuaGlu}, is one step forward to the use
of $f_{1}$ for the matter fields (see eq.\ (\ref{cellintg})),
and $f_{1,\mu}$ for the gauge fields. This already provides
a promising degree of locality.

To explore further possibilities, the obvious next step
is to consider higher orders in this pattern,
\begin{equation}
\phi_{x} = \int d^{d}y \ \Big[ \prod_{\nu =1}^{d}
f_{N} (x_{\nu}-y_{\nu}) \Big] \varphi (y) .
\end{equation}
In momentum space we find inductively
\begin{equation}
f_{N}(k) = \prod_{\mu =1}^{d} \Big( \frac{\hat k_{\mu}}{k_{\mu}}
\Big)^{N},
\end{equation}
and the perfect scalar propagator takes the form
\begin{equation}
G_{N}(k) = \sum_{l \in \Z^{d}}
\frac{\Pi ^{2N}(k+2\pi l)}{(k+2\pi l)^{2}+m^{2}} + \alpha_{N} .
\end{equation}
For increasing $N$ the peak of $f_{N}(k)$ around $k=0$
becomes steeper, in agreement with the observation that
$f_{N}(x)$ turns less localized. Still the 1d propagator
can be made ultralocal for any order $N$, if we choose
the suitable smearing parameter $\alpha_{N}$. The first
few orders require at $m=0$
\begin{equation}
\alpha_{0} = 0 \ , \ \alpha_{1} = \frac{1}{6} \ ,
\ \alpha_{2} = \frac{1}{4} - \frac{\hat k^{2}}{120} \ , \
\alpha_{3} = \frac{1}{3} - \frac{\hat k^{2}}{40} +
\frac{\hat k^{4}}{5040} \quad {\rm etc.}
\end{equation}
The general form is a polynomial in $\hat k^{2}$ of order $N-1$.\\

The use of these B-spline functions as they stand makes the
action less local for $N$ increasing beyond 1, so that we cannot find
anything better than $N=1$ (the case we had before)
within this scheme. The case $N=2$ is not that bad, but beyond
the level of locality decreases rapidly.
This suggests that a strong overlap of the blocking functions
for different sites is unfavorable for locality, which was also confirmed
by considering other types of functions.

However, we may squeeze the functions again
into the interval $[-1/2,1/2]$ on each axis by just rescaling
$x$ and  correcting the normalization,
$F_{N}(x) = N f_{N}(Nx)$.
Thus we give up the property (\ref{demo}). Instead
of a democratic treatment of the continuum points we do
something between that and decimation. 
It turns out that now the locality is in business, at least for
\begin{equation} \label{eiffel}
F_{2}(x) = 2 (1-2 \vert x \vert) \ \Theta (1/2 - \vert x \vert ) \ .
\end{equation}
We summarize these observations in Table \ref{tabBS}.
\begin{table}
\begin{center}
\begin{tabular}{|c|c|c|c|}
\hline
decay coefficient & $f_{1}$ & $f_{2}$ & $F_{2}$ \\
\hline
$c_{1}$ & 3.805 & 2.996 & 3.348 \\
\hline
$c_{2}/\sqrt{2}$ & 3.876 & 3.159 & 3.566 \\
\hline
$c_{3}/\sqrt{3}$ & 3.397 & 3.156 & 3.757 \\
\hline
$c_{4}/2$ & 3.258 & 3.179 & 3.333 \\
\hline
\end{tabular}
\end{center}
\vspace{-1mm}
\caption{\it{The competition for locality: we give the coefficients
of the exponential decay, $\vert \rho (i,0,0,0) \vert \propto
\exp \{ - c_{1} i\}, \dots , \vert \rho (i,i,i,i)\vert \propto
\exp \{ - c_{4} i\}$, fitted in the interval $i=1 \dots 5$.
The decay coefficients $c_{i}$ are divided by $\protect\sqrt{i}$, 
so that the level of rotation invariance is revealed too.
In that respect, the use of $F_{2}$, and especially $f_{2}$,
is even better than the standard blocking function $f_{1}$ 
(piece-wise constant).}}
\vspace{-2mm}
\label{tabBS}
\end{table}
We recommend to focus on $N=2$, and we denote the corresponding
blocking function $F_{2}$ as ``{\em Eiffel tower function}''.
It has the virtue that it can be related to a relatively simple
blocking from a fine to a coarse lattice, such that we reproduce
the same fixed point. If $N$ increases, it becomes more and more 
difficult to establish such relations; the cells, where the block
variables are build, have to be larger and larger. However,
this property is needed if we ultimately want to combine
the blocking from the continuum with a subsequent multigrid
minimization. Moreover, the step beyond $N=2$ is not profitable
in view of locality.

We truncate the FPA obtained from blocking with the
Eiffel tower function $F_{2}$ again to a unit hypercube.
The couplings of this new hypercube scalar are given in 
Table \ref{eif}. Its properties regarding the spectrum and
thermodynamics are on the same level as the hypercube scalar 
obtained from the standard approach and given in Table \ref{copvol};
all in all the hypercube scalar of Section 3 is slightly superior,
but in specific respects the Eiffel tower hypercube scalar is better.
\begin{table}
\begin{center}
\begin{tabular}{|c|c|c|c|c|}
\hline
$\rho (0000)$ & $\rho (1000)$ & $\rho (1100)$ & $\rho (1110)$ & $\rho (1111)$ \\
\hline
4.09671070 & -0.23791704 & -0.056450784 & -0.021127525 & -0.010154672 \\
\hline
\end{tabular}
\end{center}
\vspace{-1mm}
\caption{\it{The couplings $\rho (r)$ of the truncated perfect scalar
at $m=0$, constructed from the ``Eiffel tower'' blocking.}}
\vspace{-2mm}
\label{eif}
\end{table}

For the Wilson-like fermions we observed a qualitatively similar
behavior; in that case the Eiffel tower blocking looks even a little
better than the blocking with the usual step function, see Appendix B.

\section{Conclusions and outlook}

We have constructed perfect lattice formulations for free
scalar fields of any mass, and we optimized their locality.
We also explored new aspects of the RGT improvement technique,
in particular regarding non-hypercubic lattices and blocking 
functions, which are not piece-wise constant.
We then truncated the couplings of a perfect scalar
to a short range --  so that the formulation become applicable in 
simulations -- and showed that a drastic improvement persists.
This was observed from the dispersion relation as well as
thermodynamic properties. In view of the numerical treatment
we showed that a new variant of preconditioning is applicable
and powerful, which is also true for truncated perfect fermions.

Simulations of truncated perfect scalars have been performed
in $d=1$ for the anharmonic oscillator \cite{BS}. The 2d
truncated perfect scalars of Section 3 have been used by W.\ Loinaz
in an ongoing numerical study of the critical coupling
in the $\lambda \phi^{4}$ model, along the lines of 
Ref.\ \cite{Loinaz}. Improved scalar actions play a r\^{o}le
in bosonic spin models, in particular the non-linear
$\sigma$-model \cite{Has,Anton,2dO3}.
A potential field of application in $d=4$ is the Higgs model
(for a very recent application of Symanzik improvement, 
see Ref.\ \cite{Higgs}).
Finally the truncated perfect Laplacian could be applied
also for an improved gauge fixing in the spirit of Ref.\ \cite{ig}.\\



{\bf Acknowledgment} {\it I would like to thank Ph.\ de Forcrand for
encouraging me to finish this paper, a draft of which has
been around for more than two years.
The first part is based on techniques worked out in collaboration
with R.\ Brower, S.\ Chandrasekharan and in particular U.-J.\ Wiese.}

\appendix

\section{The perfect propagator under a RGT}

We want to demonstrate the perfectness of the propagator
$G(k)$ in eq.\ (\ref{perfprop}) by explicitly applying a
block factor $n$ RGT. The perfect action is supposed to reproduce
itself, up to a rescaling of the mass.

From Ref.\ \cite{BeWi} we can extract
the general recursion relation for the propagator
under a block factor $n$ RGT (the $n \to \infty$ limit
of which is given in eq.\ (\ref{RGT})),
\begin{equation}
G'(k)\vert_{m} = \frac{1}{n^{2}} \sum_{\bar l}
G((k+2\pi \bar l )/n)\vert_{m} \prod_{\mu =1}^{d}
\Big( \frac{\sin (k_{\mu}/2)}{n \sin ( 
(k_{\mu}+ 2\pi \bar l_{\mu})/2n)} \Big)^{2} + \alpha_{n} ,
\end{equation}
where $\bar l \in \{0,1,2,\dots ,n-1\}^{d}$. Inserting
now the perfect propagator of eq.\ (\ref{perfprop})
at mass $m$ we obtain
\begin{eqnarray}
G'(k)\vert_{m} &=& \frac{1}{n^{2}} \sum_{\bar l} \Big[ 
\frac{\alpha_{n}}{1-1/n^{2}}
\prod_{\mu} \Big( \frac{\sin (k_{\mu}/2)}{n \sin
((k_{\mu}+2\pi \bar l_{\mu})/2n)} \Big)^{2} + \nonumber \\
&& \sum_{l\in \Z^{d}} \frac{1}{(k+2\pi \bar l 
+ 2\pi nl)^{2}/n^{2}+m^{2}}
\times \nonumber \\ &&
\prod_{\mu} \Big( \frac{2 \sin ((k_{\mu}+2\pi \bar l_{\mu})/2n)}
{(k_{\mu}+2\pi \bar l_{\mu} + 2\pi nl_{\mu})/n} \
\frac{\sin(k_{\mu}/2)}{n \sin ((k_{\mu}+2\pi \bar l_{\mu})/2n)}
\Big)^{2} \Big] + \alpha_{n} \nonumber \\
&=& \alpha_{n} \Big[ 1 + \frac{1}{(n^{2}-1)n^{2d}} \sum_{\bar l}
\prod_{\mu} \Big( \frac{\sin (k_{\mu}/2)}{\sin ((k_{\mu}+
2\pi \bar l_{\mu})/2n)} \Big)^{2} \Big] + \nonumber \\
&& \sum_{\bar l} \sum_{l \in \Z^{d}} \frac{1}{(k+2\pi [\bar l
+nl])^{2} + (nm)^{2}} \Pi (k + 2\pi [\bar l + nl])^{2} \nonumber \\
&=& G(k)\vert_{nm}
\end{eqnarray}
This is the desired result, confirming the claims of
Section 2. In the last step we have used the identity
\begin{displaymath}
\sum_{j=0}^{n-1} \Big( \frac{\sin (x/2)}{\sin ((x+2\pi j)/2n)}
\Big)^{2} \equiv n^{2} .
\end{displaymath}
Amazingly, I could not find this -- or a directly
related -- identity in any
table of formulae. It can be demonstrated, however,
by inserting
\begin{displaymath}
\frac{\sin (ny/2)}{\sin (y/2)} \equiv \sum_{k=-(n-1)/2}^{(n-1)/2}
\exp (i k y)
\end{displaymath}
(for odd $n$ the sum runs over half integers)
and then summing over $j$.

\section{Eiffel tower blocking for fermions}

It is straightforward to extend the considerations for non hypercubic
lattices and non piece-wise constant blocking functions (Sections
6 and 7) to fermions. The most promising variant for applications
is the Eiffel tower blocking, so we add in this appendix the
corresponding results for Wilson-type fermions. We denote the
lattice Dirac operator as $D = \rho_{\mu} \gamma_{\mu} + \lambda$,
where $\rho_{\mu}$ is odd in $\mu$ direction and even otherwise,
while the Dirac scalar $\lambda$ is entirely even.
The prefect free propagator obtained from Eiffel tower blocking
-- i.e.\ using the blocking function $F_{2}$ given in 
eq.\ (\ref{eiffel}) -- reads
\begin{equation}
D^{-1}(k) = \sum_{l \in Z \!\!\! Z ^{d}} 
\frac{1}{i (\sla k + 2\pi \sla l ) +m}
\prod_{\mu =1}^{d} \left( \frac{8 [1-\cos((k_{\mu}+2\pi l_{\mu})/2)]}
{(k_{\mu}+2\pi l_{\mu})^{2}} \right) ^{2} + \alpha_{f} .
\end{equation}
For massless fermions, 1d ultralocality is provided by 
the RGT smearing parameter $\alpha_{f} = 1/2$ 
(this is exactly the same value which is also required for the piece-wise
constant blocking function $f_{1}=F_{1}$ \cite{QuaGlu}). Here even $F_{N>2}$
yields comparable locality of $D$, but for reasons pointed out in Section 7 
we prefer to stay with $F_{2}$. There the locality of the FPA
is slightly better than for $f_{1}$: using $F_{2}$, the decay 
along the  4d diagonal has the exponential coefficients (analogous to $c_{4}$
in Section 2) $5.037$ (for $\rho_{\mu}$) and $5.109$ (for $\lambda$),
to be compared with  $4.931$ (for $\rho_{\mu}$) and $5.039$ (for $\lambda$)
when we use $f_{1}$. Therefore truncation yields a hypercube fermion, which
is at least of the same quality as the one presented in Ref.\ \cite{StL};
its dispersion relation does not start with an overshoot, it is
excellent up to $\vert \vec k \vert \approx 2$, and also other properties 
are in business. The new hypercube fermion couplings are given
in Table \ref{tabHF}.

\begin{table}
\begin{center}
\begin{tabular}{|c|c|c|}
\hline
$y-x$ & $\rho_{1}(y-x)$ & $\lambda (y-x)$ \\
\hline
$ (0,0,0,0)$ & 0 & 1.83043692 \\
\hline
$ (1,0,0,0)$ & 0.158134654 & $-0.0577452799$ \\
\hline
$ (1,1,0,0)$ & 0.0290077129 & $-0.0295719810$ \\
\hline
$(1,1,1,0)$ & 0.0108088580 & $-0.0161376837$ \\
\hline
$(1,1,1,1)$ & 0.00476409657 & $-0.00889632880$ \\
\hline
\end{tabular}
\end{center}
\caption{\it{The couplings of the massless hypercube fermion constructed
by using the Eiffel tower function $F_{2}$ as RGT blocking function,
and by truncating the FPA. }}
\label{tabHF}
\end{table}

\end{document}